\documentclass[twocolumn,groupedaddress,assymb]{revtex4}
\usepackage{graphicx}
\usepackage{amssymb}
\usepackage{amsmath}

\begin{document}

\title{Weak measurement amplification based on thermal noise effect}
\author{Gang Li\footnote{ligang0311@sina.cn},$^{1}$ Tao Wang\footnote{suiyueqiaoqiao@163.com},$^{2}$ and He-Shan Song$^{3}$}
\affiliation{$^{1}$School of Physics and Electronic Information, Yan'an University, Yanan
716000, China}
\affiliation{$^{2}$College of Physics, Tonghua Normal University, Tonghua 134000, China}
\affiliation{$^{3}$School of Physics, Dalian University Of Technology, Dalian 116024,
China}
\date{\today }

\begin{abstract}
Most studies for postselected weak measurement focus on using pure Gaussian
state as a pointer, which can only give an amplification limit reaching the
level of the ground state fluctuation. When the pointer is initialised in a
thermal state, we find that the amplification limit after the postselection
can reach the level of thermal fluctuation, indicating that the
amplification effect achieving the level of thermal fluctuation is also
increased with the temperature grow, and also give the amplification
mechanism different from the one with pure Gaussian state pointer. To
illustrate these results, we propose two schemes to implement room
temperature amplification of the mechanical oscillator's displacement caused
by a single photon in optomechanical system. The two schemes can both
enhance the mechanical oscillator's original displacement by nearly seven
orders of magnitude, attaining sensitivity to displacements of $\thicksim $ $%
0.26$ nm. Such amplification effect can be used to observe the impact of a
single photon on a room temperature mechanical oscillator which is hard to
detect in traditional measurement. ~~~~~\newline
~~~~~\newline
\end{abstract}

\maketitle

\section{Introduction}

Weak measurement (WM) with postselection, first proposed by Aharonov et al.
\cite{Aharonov88}, is an enhanced detection scheme where the system is
weakly coupled to the pointer. The postselection on the system leads to an
unusual effect: the average displacement of the postselection pointer is far
beyond the the eigenvalue spectrum of the system observable, in contrast to
von Neumann measurement. The mechanism behind this effect is the
superposition (interference) between different postselection pointer states
\cite{Duck89}. Much theoretical research based of weak value is shown in
\cite{Jozsa07,Shikano10,Turek15}. WM has been realized \cite{Ritchie91}, and
proven applicable to amplify tiny physical effects \cite%
{Hosten08,Dixon09,Gorodetski12,Starling09,Starling10}. More experimental
protocols have been proposed \cite{proposal-electron
spin,proposal-Tomography of Many-Body Weak
Value,proposal-phaseshift,proposal-chargesensing,proposal-nonlinear
crystal,proposal-time delay of
light,proposal-entangled1,proposal-entangled2,proposal-Justin-recycled
photons}. A Fock-state view for WM is given in \cite{Simon11}, based on
which a WM protocols combined with optomechanical system \cite%
{Marquardt09,Aspelmeyer14} is proposed \cite{Li14,Li--15,Li15,Pepper12}, and
more applications of the field are reviewed in \cite{Kofman12,Dresse14}.

In most previous studies the pointer is initialised in pure Gaussian state.
It was an inherit assumption that the pointer has to be in the pure state at
the inception of WM \cite{Aharonov88,Duck89}. A pointer can be easily
represented with light in pure state \cite{Ritchie91,Hosten08,Dixon09}, but
with particles of efficient mass \cite%
{Kleckner01,Arndt99,Hackermuller04,Riou06}, it's difficult to initialize
them in pure state due to environmental induced decoherence. Recently using
squeezed pointer states combined with WM can also amplify small physical
quantities \cite{Li--15}. Moreover, weak measurement based of thermal state
pointer can also enhance parameter estimation in quantum metrology\
discussed in \cite{Li-15} which is very different from previous results \cite%
{Tanaka13,Knee14,Zhang15,Knee15}. The discussion of mixed state pointer in
WM is given in \cite{Johansen04,Cho10,Tamate12}. However, they only focus on
weak-value formalism (see \cite{Kofman12,Dresse14} for reviews) but not what
extent the amplification value can be, i.e., the amplification limit.
Needless to say, thermal state is easier to prepare, especially in
optomechanical systems \cite{Li14,Pepper12}. One may naturally ask whether
using thermal state pointer in WM can give a valid result for the
amplification limit, and what kind of advantage it has than pure state
pointer.

In the paper, we study the limits of amplifying tiny physical quantities or
effects based on weak measurement. Our paper begins with a general
discussion about weak measurement with a thermal state pointer, and show
that the maximal displacement of the postselection pointer, proportional to
the imaginary weak value, can reach the level of thermal fluctuation, which
is much larger than the ground state fluctuation with pure state pointer
\cite{Simon11,Li14}. As the temperature grows, the amplification effect
achieving the level of thermal fluctuation is also increased, thereby
constantly improving the amplification limit, indicating that thermal noise
effect of the pointer is beneficial for weak measurement amplification. This
amplification is attributed to two probabilistic average results: one is the
classical statistical properties of thermal state itself, and the other is
the representation of quantum statistical probability, namely, the
superposition of the number state $|n\rangle $ and the state $(c+c^{\dag
})|n\rangle $ (unnormalized) of the postselection pointer. Such
superposition is the generalization of the mechanism behind the
amplification in Ref. \cite{Simon11,Li14,Li--15,Li15}.

We apply the general idea to the field of optomechanical system. We find
that the amplification of the mirror's displacement occurred at time near
zero is very important for bad cavities with non-sideband resolved regime,
and can overcome the shortcomings of difficultly observing the amplification
effect due to dissipation \cite{Li14}. Finally we show that the unique
advantage of our schemes is that the amplification at room temperature, with
current experimental technologies, can be used to observe the impact of a
single photon on a room temperature mechanical oscillator which is hard to
detect in traditional measurement \cite{Temperature effect}.

The structure of our paper is as follows. In Secs. II, we give a general
discussion about weak measurement with a thermal state pointer. In Secs.
III, we state the second main result of this work, including weak
measurement amplification in optomechanical system using phase shifter $%
\theta $ and using a displaced thermal state, respectively. In Secs. IV, we
give the conclusion about the work, respectively.

\section{Fock-state view of weak measurement with a thermal state pointer}

In the standard scenario of WM, the interaction Hamiltonian between the
system and the pointer is $\hat{H}=\chi (t)\hat{A}\hat{q}$ (setting $\hbar
=1 $), where $A$ is a system observable, $q$ is the position observable of
the pointer and $\chi (t)$ is a narrow pulse function with interaction
strength $\chi $. As in Ref. \cite{Simon11}, if we define $\hat{c}=\hat{q}%
/2\sigma +i\sigma \hat{p}$, the interaction Hamiltonian can be rewritten as
\begin{equation}
\hat{H}=\chi (t)\sigma \hat{A}(\hat{c}+\hat{c}^{\dagger }),  \label{I}
\end{equation}%
where $\hat{q}=$ $\sigma (\hat{c}+\hat{c}^{\dagger })$, $\hat{p}%
=-i(c-c^{\dagger })/(2\sigma )$, and $\sigma $ is the zero-point
fluctuation. Suppose the initial system state is $|\psi _{i}\rangle
=(|a_{1}\rangle _{s}+|a_{2}\rangle _{s})/\sqrt{2}$, where $a_{1}$ and $a_{2}$
is eigenvalues of $A$. Then we consider the initial pointer state as
\begin{equation}
\rho _{th}(z)=(1-z)\sum_{n=0}z^{n}|n\rangle _{m}\langle n|_{m},  \label{II}
\end{equation}%
with $z=e^{-\beta \omega _{m}}$ and $\beta =1/(k_{B}T)$, where $k_{B}$ is
the Boltzmann constant and $T$ is the temperature.

Next we make a postselection of the state of the measured system. Because of
the linearity of $\rho _{th}(z)$, we need only look at the component number
states $|n\rangle _{m}$ are weakly coupled with $|\psi _{i}\rangle $ using
Eq. (\ref{II}). Then we postselect the system into a final state $|\psi
_{p}\rangle =[\cos (\pi /4-\varepsilon )|a_{1}\rangle _{s}-e^{i\varphi }\sin
(\pi /4-\varepsilon )|a_{2}\rangle _{s}]$ with $\varphi \ll 1$ and $%
\varepsilon \ll 1$, which is nonorthogonal to $|\psi _{i}\rangle $, i.e., $%
\langle \psi _{p}|\psi _{i}\rangle \approx \varepsilon +i\varphi /2$, then
the reduced pointer state after the postselection for each $n$ component of
the pointer state is given by
\begin{eqnarray}
|\psi _{m}(n)\rangle &=&\langle \psi _{p}|\exp [-i\eta \hat{A}(\hat{c}+\hat{c%
}^{\dagger })]|\psi _{i}\rangle |n\rangle _{m}  \notag \\
&=&([\cos (\pi /4-\varepsilon )D(-ia_{1}\eta )-e^{-i\varphi }\sin (\pi /4
\notag \\
&-&\varepsilon )D(-ia_{2}\eta )]|n\rangle _{m})/\sqrt{2},  \label{III}
\end{eqnarray}%
where $\eta =\chi \sigma $ and $D(\alpha )=\exp [\alpha \hat{c}^{\dagger
}-\alpha ^{\ast }\hat{c}]$ is a displacement operator.

When $\varphi \ll 1$, $\varepsilon \ll 1$ and $\eta (2n+1)^{1/2}\ll 1$,
i.e., $\eta \ll 1$, the approximation of Eq. (\ref{III}) is (normalized)
\begin{eqnarray}
|\psi _{m}(n)\rangle _{\eta \ll 1} &\approx &B_{n}[(2\varepsilon +i\varphi
)|n\rangle _{m}+i\eta (a_{2}  \notag \\
&-&a_{1})(\hat{c}+\hat{c}^{\dagger })|n\rangle _{m}]/2,  \label{IV}
\end{eqnarray}%
where $B_{n}=2[4\varepsilon ^{2}+\varphi ^{2}+\eta
^{2}(a_{2}-a_{1})^{2}(2n+1)]^{-1/2}$ is a normalization coefficient for each
state $|\psi _{m}(n)\rangle _{\eta \ll 1}$, and the final total pointer
state after the postselection is (normalized)
\begin{equation}
\rho _{pm}=(1-z)\sum_{n=0}z^{n}|\psi _{m}(n)\rangle _{\eta \ll 1}\langle
\psi _{m}(n)|_{\eta \ll 1}/B_{tot},  \label{ppi}
\end{equation}%
where $B_{tot}=(\sigma ^{2}\varphi ^{2}+4\sigma ^{2}\varepsilon ^{2}+\sigma
_{q}^{2}(a_{2}-a_{1})^{2}\eta ^{2})/(4\sigma ^{2})$ is a normalized
coefficient for $\rho _{pm}$, and $\sigma _{q}=\coth ^{1/2}(\beta \omega
_{m}/2)\sigma $ represents thermal fluctuations of the position $q$ space.

Special note is given here, we only discuss the problem beyond the
weak-value amplification \cite{real amplification}, which can reach the
maximum amplification value. In addition, the discussion of the weak-value
amplification with imaginary and real values can be seen in the Appendix B,
respectively. For Eq. (\ref{IV}), when $M=q$ and $\varepsilon =0$, the
displacement of the pointer for each state $\psi _{m}(n)\rangle _{\eta \ll
1} $ is (see Appendix A)

\begin{eqnarray}
\langle q\rangle _{n} &=&B_{n}^{2}CTr(\{M,q\}|n\rangle _{m}\langle
n|_{m})/\sigma  \notag \\
&=&\sigma B_{n}^{2}C(2n+1)  \label{ppk}
\end{eqnarray}%
with $C=\varphi \eta (a_{2}-a_{1})/2$, where $\{\cdot \}$ denote
anticommutation rules in quantum mechanics and $Tr(\cdot \rho )$ as $\langle
\cdot \rangle _{\rho }$ with any state $\rho $ for short throughout the
paper. We note that $(2n+1)\sigma $ is due to the anticommutation
interaction between the superposition of the number states $|n\rangle $ and $%
(\hat{c}+\hat{c}^{\dagger })|n\rangle $ (unnormalized) and the measured
observable $M$ ($M=q$), i.e.,
\begin{eqnarray}
\langle n|\{M,q\}|n\rangle &=&\langle \psi _{m}(n)|_{\eta \ll 1}q|\psi
_{m}(n)\rangle _{\eta \ll 1}/B_{n}C  \notag \\
&=&(2n+1)\sigma .  \label{piao}
\end{eqnarray}

For Eq. (\ref{ppi}), the average displacement of the pointer in position $q$
space will be
\begin{eqnarray}
\langle \hat{q}\rangle &=&\sum_{n=0}P_{n}\langle q\rangle _{n}  \notag \\
&=&C\sigma _{q}^{2}/(\sigma B_{tot})  \label{ooi}
\end{eqnarray}%
and $\langle \hat{p}\rangle =0$, where $P_{n}=z^{n}(1-z)B_{n}^{-2}/B^{tot}$
is the classical statistical probability for each state $|\psi
_{m}(n)\rangle _{\eta \ll 1}$ in the ensemble of the pure state $\left\{
P_{n},\psi _{m}(n)\rangle _{\eta \ll 1}\right\} $. Multiplying the classical
probability $P_{n}$ and the corresponding displacement $\langle q\rangle
_{n} $, we get
\begin{equation}
P_{n}\langle q\rangle _{n}=\sigma C(1-z)z^{n}(2n+1)/B_{tot}.  \label{ooii}
\end{equation}%
We note that $(1-z)z^{n}$ in Eq. (\ref{ooii}) is due to the classical
statistical properties of thermal state itself. As the temperature $T$
grows, there is an increased occupancy of the higher number states $\left\{
|n\rangle _{m},(\hat{c}+\hat{c}^{\dagger })|n\rangle _{m}\right\} $ in
thermal pointer (\ref{ppi}). These higher number states have more energy and
so they can cause a higher displacement of the pointer than the lower number
states. Therefore, the average displacement of the pointer $\langle \hat{q}%
\rangle $ in Eq. (\ref{ooi}) is increased with the increase of the
temperature $T$.

From Eq. (\ref{ooi}), we can see that $\langle \hat{q}\rangle $ is non-zero
in position $q$ space, and get the maximal positive and negative values $\pm
\sigma _{q}$ (thermal fluctuation) when $\varphi =\pm \sigma
_{q}(a_{2}-a_{1})\eta /\sigma $, respectively, which are much larger than
that using the pure state pointer \cite{Aharonov88,Simon11,Li14}, i.e., the
ground state fluctuation $\sigma $. We find that as the temperature
increases, the maximum value $\pm \sigma _{q}$ is futher increased.
Therefore, the $|\psi _{m}(n)\rangle $ components corresponding to the
maximal positive and negative amplification are, respectively, $|\psi
_{m}(n)\rangle _{\max ,\eta \ll 1}=[\sigma _{q}/\sigma \pm (\hat{c}+\hat{c}%
^{\dagger })]|n\rangle ]/2$. Obviously, the key to understand the
amplification is attributed to two probabilistic average results: one is the
superposition of the number states $|n\rangle $ and $(\hat{c}+\hat{c}%
^{\dagger })|n\rangle $ in a thermal postselection pointer, which originate
from the representation of quantum statistical probability. This result
reveals the more generalized law of causing amplification effect since it is
regarded as a generalization of the mechanism behind the amplification in
standard WM \cite{Aharonov88,Simon11}, which is the superposition of the
ground state $|0\rangle $ and the one phonon state $|1\rangle $ of the
postselection pointer (see Appendix C); the other is the ensemble of the
pure state $\left\{ P_{n},\psi _{m}(n)\rangle _{\eta \ll 1}\right\} $, which
originated from the representation of the classical statistical properties
of thermal state itself. In a word, thermal noise effect of the pointer is
beneficial for the amplification of the displacement corresponding to the
imaginary part of weak value. It is surprised that in \cite{Li19} the
approach above can also enhance parameter estimation in quantum metrology.

\section{Weak measurement amplification of one photon in optomechanical
system}

\subsection{Optomechanical model}

To show how the above results can be applied, we consider a March-Zehnder
interferometer combined with optomechanical system where the optomechanical
cavity (OC) A and the stationary Fabry-Perot cavity B is embedded in its one
and another arm, respectively (see Fig. 1), the Hamiltonian is given by,
\begin{equation}
\hat{H}=\hbar \omega _{c}(a^{\dag }a+b^{\dag }b)+\hbar \omega _{m}c^{\dag
}c-\hbar ga^{\dag }a(c+c^{\dag }),  \label{ee}
\end{equation}%
where $\omega _{c}$ is the frequency of the optic cavity A, B of length $L$
with corresponding annihilation operators $\hat{a}$ and $\hat{b}$, $\omega
_{m}$ is the angular frequency of mechanical system with corresponding
annihilation operator $\hat{c}$, and the optomechanical coupling strength $%
g=\omega _{0}\sigma /L$, $\sigma =(\hbar /2m\omega _{m})^{1/2}$ which is the
zero point fluctuation and $m$ is the mass of mechanical system. Here it is
a weak measurement model where the mirror is used as the pointer to measure
the number of photon in cavity A, with $a^{\dag }a$ of Eq. (\ref{ee})
corresponding to $\hat{A}$ in Eq. (\ref{I}) in the standard scenario of weak
measurement (see Appendix E).
\begin{figure}[tp]
\includegraphics[width=2.3in]{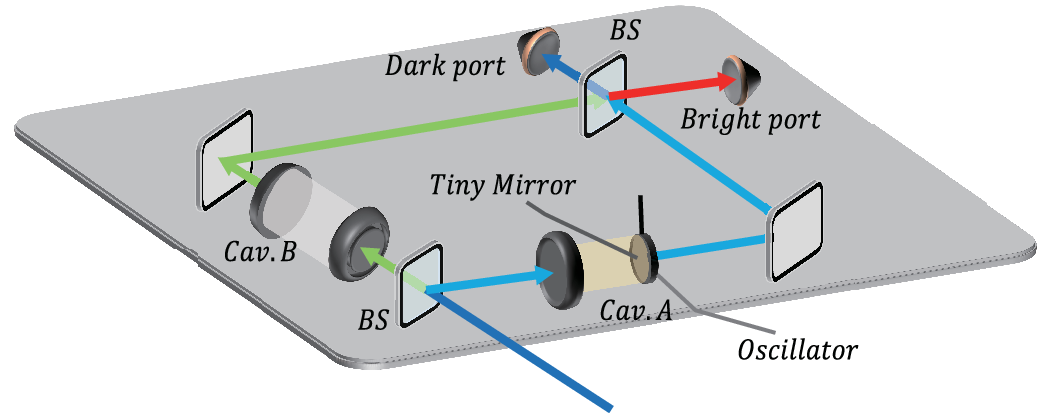}.
\caption{The photon enters the first beam splitter of March-Zehnder
interferometer, before entering an optomechanical cavity A and a
conventional cavity B. The photon weakly excites the tiny mirror. After the
second beam splitter, and dark port is detected, i.e., postselection acts on
the case where the mirror has been excited by a photon, and fails otherwise.}
\end{figure}

\subsection{Weak measurement amplification using a phase shifter $\protect%
\theta $}

As shown in Fig. 1, suppose one photon enters the interferometer, after the
first beam splitter and a phase shifter $\theta $ in the arm A of the
interferometer, the initial state of the photon becomes $|\psi _{i}(\theta
)\rangle =(1/\sqrt{2})(e^{i\theta }|1\rangle _{A}|0\rangle _{B}+|0\rangle
_{A}|1\rangle _{B})$ with $\theta \ll 1$. The mirror is initialised in
thermal state $\rho _{th}(z)$. After a weak interaction using Eq. (\ref{ee}%
), according to the results of the Hamiltonian in \cite{Bose97,Mancini97},
the state of the total system will be%
\begin{eqnarray}
\rho (z) &=&(1-z)\sum_{n=0}z^{n}[|1\rangle _{A}|0\rangle _{B}e^{i(\phi
(t)+\theta )}D(\xi )  \notag \\
&+&|0\rangle _{A}|1\rangle _{B}]|n\rangle _{m}\langle n|_{m}[\langle
1|_{A}\langle 0|_{B}e^{-i(\phi (t)+\theta )}  \notag \\
&&D^{\dagger }(\xi )+\langle 0|_{A}\langle 1|_{B}]/2,  \label{cc}
\end{eqnarray}%
where $\xi (t)=k(1-e^{-i\omega _{m}t})$ and $\phi (t)=k^{2}(\omega
_{m}t-\sin \omega _{m}t)$ with $k=g/\omega _{m}$. Then the second beam
splitter postselects for the photon state $|\psi _{p}\rangle =(|1\rangle
_{A}|0\rangle _{B}-|0\rangle _{A}|1\rangle _{B})/2$, which is nonorthogonal
to $|\psi _{i}(\theta )\rangle $, i.e., $\langle \psi _{p}|\psi _{i}(\theta
)\rangle \approx i\theta /2$ (imaginary), in other words, when a photon is
detected at the dark port, the reduced state of the mirror after the
postselection becomes (see Appendix D, unnormalized)

\begin{equation}
\rho _{m}^{pha}=(1-z)\sum_{n=0}z^{n}|\psi _{1}(n)\rangle \langle \psi
_{1}(n)|,  \label{dd}
\end{equation}%
where $|\psi _{1}(n)\rangle =[e^{i(\phi (t)+\theta )}D(\xi )|n\rangle
_{m}-|n\rangle _{m}]/2$ denotes the $n$ phonon component state of the mirror.

Substituting (\ref{dd}) into the displacement expression of the pointer (\ref%
{eq:1-5}) in Appendix A, and applying the identity of the associated
Laguerre polynomial $L_{n}^{k}(x)$ \cite{Gradshteyn65},

\begin{equation}
\sum_{n=0}^{\infty }L_{n}^{k}(x)z^{n}=(1-z)^{-k-1}\exp [-xz/(1-z)],
\label{oo}
\end{equation}%
the average displacement $\langle q(t)\rangle $ of the mirror over all $n$
phonon component states $|\psi _{1}(n)\rangle $ is (see Appendix B for
detail derivation)
\begin{eqnarray}
\langle q(t)\rangle &=&\sigma \lbrack \xi +\xi ^{\ast }-(1-z)^{-1}(\Phi \xi
+\Phi ^{\ast }\xi ^{\ast }  \notag \\
&-&z[\Phi \xi ^{\ast }+\Phi ^{\ast }\xi ])]/(2-\Phi -\Phi ^{\ast }),
\label{gg}
\end{eqnarray}%
where $\Phi =\exp (-\sigma _{q}|\xi |^{2}/(2\sigma )+i\phi (t)+i\Omega )$
with $\Omega =\theta $.

Figure 2(a) show that the average displacement $\langle q(t)\rangle /\sigma $
of the mirror versus time $\omega _{m}t$. At time near $\omega _{m}t=0$ the
maximal amplification can reach $\sigma _{q}$ (thermal fluctuation) which is
$\sqrt{19}\sigma $ when $z=0.9$. This result is beyond the strong-coupling
limiting $\sigma $ (the ground-state fluctuation) \cite{Marshall03}.
Therefore, thermal noise effect of the mirror is beneficial for the
amplification of the mirror's displacement caused by one photon, which means
that the impact of one photon on an mechanical oscillator with arbitrary
temperature can be observed.

In order to observe the amplification effects appearing at time near $T=0$,
for Eq. (\ref{dd}) we can then perform a small quantity expansion about time
$T$ till the second order. Suppose that $|\omega _{m}t-T|\ll 1$, i.e., $%
\omega _{m}t\ll 1$, $k\ll 1$ and $\theta \ll 1$, then the approximation of $%
|\psi _{1}(n)\rangle $ is given by (normalized)
\begin{equation}
\psi _{1}(n)\rangle _{\omega _{m}t\ll 1}=B_{1}(n)[i\theta +ik\omega
_{m}t(c+c^{\dagger }))]|n\rangle /2,  \label{hh}
\end{equation}%
where $B_{1}(n)=2[\theta ^{2}+k^{2}(\omega _{m}t)^{2}(2n+1)]^{-1/2}$ is a
normalization coefficient for each state $|\psi _{1}(n)\rangle _{\omega
_{m}t\ll 1}$. Note that in the ensemble of pure state $\left\{ P_{n},\psi
_{1}(n)\rangle _{\omega _{m}t\ll 1}\right\} $ the classical statistical
probability for each state $|\psi _{1}(n)\rangle _{\omega _{m}t\ll 1}$ is $%
P_{n}=z^{n}(1-z)B_{1}^{-2}(n)/B_{1}^{tot}$, where $B_{1}^{tot}=(\theta
^{2}\sigma ^{2}+k^{2}(\omega _{m}t)^{2}\sigma _{q}^{2})/(4\sigma ^{2})$ is a
normalized coefficient for $\rho _{m}^{pha}$ in Eq. (\ref{hh}).
\begin{figure}[tp]
\centering
\includegraphics[width=0.45\columnwidth]{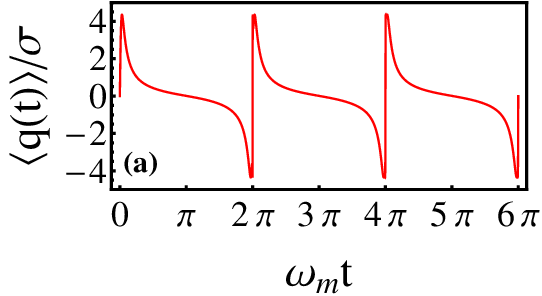} %
\includegraphics[width=0.45\columnwidth]{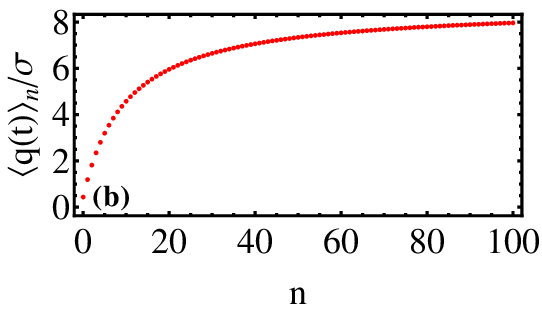}
\caption{(a) Average displacement $\langle q(t)\rangle /\protect\sigma $
versus time $\protect\omega _{m}t$ with $\protect\theta =0.0005$, $k=0.005$
and $z=0.9$. (b) Average displacement $\langle q(t)\rangle _{n}/\protect%
\sigma $ as function of $n$ when $\protect\theta =(\frac{1+z}{1-z})^{1/2}k%
\protect\omega _{m}t$ with $\protect\theta =0.0005$, $k=0.005$ and $z=0.9$.}
\end{figure}

For Eq. (\ref{hh}), the displacement $\langle q(t)\rangle _{n}/\sigma $ for
each $n$ phonon component state $|\psi _{1}(n)\rangle _{\omega _{m}t\ll 1}$
(see Appendix C) is%
\begin{equation}
\langle q(t)\rangle _{n}=B_{1}^{2}(n)\sigma \theta k\omega _{m}t(2n+1)/2.
\label{ff}
\end{equation}

In Fig. 2(b), we plot the displacement $\langle q(t)\rangle _{n}/\sigma $
for $|\psi _{1}(n)\rangle _{\omega _{m}t\ll 1}$ as function of $n$ when $%
\theta =\sigma _{q}k\omega _{m}t/\sigma $. This condition is to make achieve
the maximal value. It shows the amplification values grow with the increase
of $n$. Obviously, the superposition of $|n\rangle $ and $(c+c^{\dagger
})|n\rangle $ is the key to obtain amplification at time near $\omega
_{m}t=0 $. Note that suppose the initial pointer state is $|n\rangle _{m}$,
for the the displacement $\langle q(t)\rangle _{n}/\sigma $ in Eq. (\ref{ff}%
) we can see that the maximal amplification can reach $\pm
(2n+1)^{1/2}\sigma $ (thermal fluctuation) if $\theta =k\omega
_{m}t(2n+1)^{1/2}$, and its amplification value tends to $\infty $ with the
increase of $n$. This is different from the result in fig. 2(b). Summing the
displacement $\langle q(t)\rangle _{n}/\sigma $ for all $n$ phonon component
states $|\psi _{1}(n)\rangle _{\omega _{m}t\ll 1}$, the maximal values of
the average displacement are $\langle q(t)\rangle /\sigma =\pm \sigma
_{q}/\sigma $ (thermal fluctuation) when $\theta =\pm \sigma _{q}k\omega
_{m}t/\sigma $, respectively, and as the temperature $T$ grows, the maximal
values $\pm \sigma _{q}$ is also increased.

\subsection{Weak measurement amplification using a displaced thermal state}

Besides the above amplification scheme, as shown in Fig. 1, we can also
provide an alternative where the mirror is initialised in the displaced
thermal state \cite{Sairo96} using classical light pulses drive, $\rho
_{th}(z,\alpha )=D(\alpha )\rho _{th}(z)D^{\dagger }(\alpha )$. Without the
phase shifter $\theta $, the initial state of the photon after the first
beam splitter is $|\psi _{i}\rangle =(|1\rangle _{A}|0\rangle _{B}+|0\rangle
_{A}|1\rangle _{B})/\sqrt{2}$. Similar to the previous scheme in which weak
measurement amplification can be performed by using a phase shifter $\theta $%
, when a photon is detected at the dark port, the reduced state of the
mirror after the orthogonal postselection (i.e., $\langle \psi _{p}|\psi
_{i}\rangle =0$) is given by (see Appendix D, unnormalized)

\begin{equation}
\rho _{m}^{dis}=(1-z)\sum_{n=0}z^{n}|\psi _{2}(n)\rangle \langle \psi
_{2}(n)|,  \label{aaa}
\end{equation}%
where $|\psi _{2}(n)\rangle =[e^{i(\phi (t)+\phi (\alpha ,t))}D(\xi
(t))|n\rangle -|n\rangle ]/2$ denotes the $n$ phonon component state of the
mirror and $\phi (\alpha ,t)=-i[\alpha \xi (t)-\alpha ^{\ast }\xi ^{\ast
}(t)]$ is caused by noncommutativity of quantum mechanics \cite{Li15}.
Similar to the previous section, when substitute (\ref{aaa}) into (\ref%
{eq:1-5}) in Appendix A, the expression for the average displacement $%
\langle q(t)\rangle $ of the mirror for $\rho _{m}^{dis}$ is similar to Eq. (%
\ref{gg}) (see Appendix D for detail derivation), just with $\phi (\alpha
,t) $ instead of $\theta $.

Figure 3(a) shows that the average displacement $\langle q(t)\rangle /\sigma
$ of the mirror versus time $\omega _{m}t$. Obviously, at time near $\omega
_{m}t=0$, the maximal amplification can reach $\sigma _{q}$ (thermal
fluctuation) which is $\sqrt{19}\sigma $ when $z=0.9$. The meaning of this
result is the same as the one using a phase shifter and a huge impact of a
single photon on a high temperature mechanical oscillator can be observed.
\begin{figure}[bp]
\includegraphics[width=0.45\columnwidth]{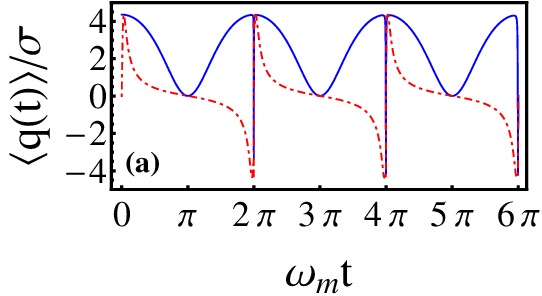} %
\includegraphics[width=0.45\columnwidth]{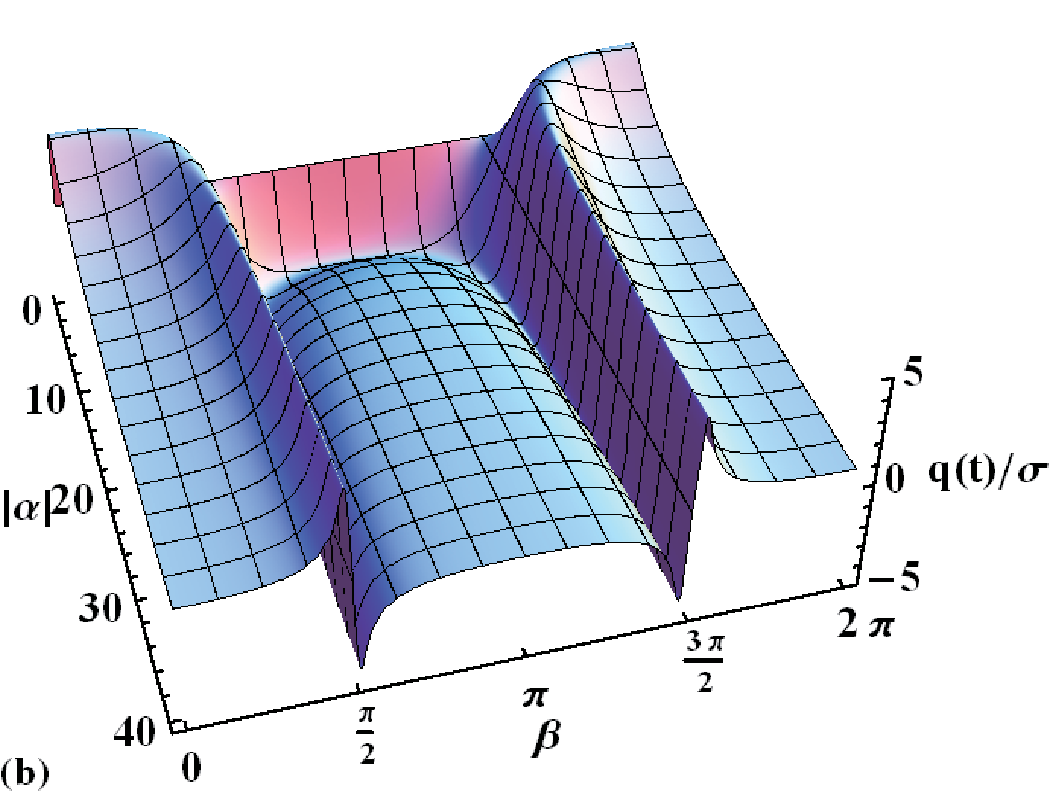}
\caption{(a) Average displacement $\langle q(t)\rangle /\protect\sigma $
versus time $\protect\omega _{m}t$ for $|\protect\alpha |=(\frac{1+z}{1-z}%
)^{1/2}/2$, $\protect\beta =0$ (blue line) and $|\protect\alpha |=10(\frac{%
1+z}{1-z})^{1/2}$, $\protect\beta =\protect\pi /2$ (red line). (b) Average
displacement $\langle q(t)\rangle /\protect\sigma $ at time $\protect\omega %
_{m}t=0.001$ as a function of $\protect\alpha =|\protect\alpha |e^{i\protect%
\beta }$; other parameters are the same as before, i.e., $k=0.005$ and $%
z=0.9 $ }
\end{figure}

Similar to Eq. (\ref{hh}), the approximation of $|\psi _{2}(n)\rangle $ is
(see Appendix D, normalized)
\begin{eqnarray}
|\psi _{2}(n)\rangle _{\omega _{m}t\ll 1} &=&B_{2}(n)[i2k|\alpha |\zeta
|n\rangle +ik\omega _{m}t(c  \notag \\
&+&c^{\dagger })|n\rangle ]/2  \label{bb}
\end{eqnarray}%
when $k\ll 1$ and $2k|\alpha |\zeta \ll 1$, where $\zeta =[(\omega
_{m}t)^{2}\sin \beta ]/2+\omega _{m}t\cos \beta $ and $B_{2}(n)=2[4k^{2}|%
\alpha |^{2}\zeta ^{2}+k^{2}(\omega _{m}t)^{2}(2n+1)]^{-1/2}$ is a
normalization coefficient for each state $|\psi _{2}(n)\rangle _{\omega
_{m}t\ll 1}$. Note that in the ensemble of pure state $\left\{ P_{n},\psi
_{2}(n)\rangle _{\omega _{m}t\ll 1}\right\} $ the classical statistical
probability for each state $|\psi _{2}(n)\rangle _{\omega _{m}t\ll 1}$ is $%
P_{n}=z^{n}(1-z)B_{2}^{-2}(n)/B_{2}^{tot}$, where $B_{2}^{tot}=(4k^{2}|%
\alpha |^{2}\zeta |^{2}\sigma ^{2}+k^{2}(\omega _{m}t)^{2}\sigma
_{q}^{2})/(4\sigma ^{2})$ is a normalized coefficient for $\rho _{m}^{dis}$
in Eq. (\ref{aaa}). This indicates that the superposition of $|n\rangle $
and $(c+c^{\dagger })|n\rangle $ is the key to obtain amplification at time
near $\omega _{m}t=0$. Fig. 3(b) show that at time $\omega _{m}t$ $=0.001$,
the average displacement $\langle q(t)\rangle /\sigma $ of the mirror as a
function of $\alpha =|\alpha |e^{i\beta }$, i.e., different displaced
thermal states $\rho _{th}(z,\alpha )$.

\subsection{Dissipation}

When the mirror is considered in a thermal bath characterized by a damping
constant $\gamma _{m}$, we have
\begin{eqnarray}
d\rho (t)/dt &=&-i[H,\rho (t)]/\hbar +\gamma _{m}\mathcal{D}[c]/(1-z)  \notag
\\
&+&\gamma _{m}z\mathcal{D}[c^{\dag }]/(1-z),  \label{ccc}
\end{eqnarray}%
where $\mathcal{D}[o]=o\rho (t)o^{\dag }-o^{\dag }o\rho (t)/2-\rho
(t)o^{\dag }o/2$. In Fig. 4(a) and Fig. 5(a), we show that at time $t\ll 1$,
the average displacements of the mirror (see Appendix E) from the exact
solution of Eq. (\ref{ccc}) for the first and the second proposed schemes,
respectively. They show that at room temperature $300K$, even if the damping
coefficient $\gamma $ ($\gamma =\gamma _{m}/\omega _{m}$) become very large,
such as $\gamma =50$, the average displacement of the mirror is the same as
the one without dissipation, $\gamma =0$, but actually the damping
coefficient of the OC we use in \cite{Pepper12} is $5\times 10^{-7}$, which
is no effect on the amplification.

\subsection{Experimental requirements}

First, we discuss the photon arrival rate versus time. Suppose a single
photon in short-pulse limit enters to the cavity. The probability density of
a photon being released from OC after time $t$ is $\kappa \exp (-\kappa t)$,
with $\kappa $ being cavity decay rate. The successful postselection
probability being released after $t$ is $[2-\exp [-\sigma _{q}^{2}|\xi
(t)|^{2}/(2\sigma ^{2})](e^{i(\phi (t)+\Omega )}+e^{-i(\phi (t)+\Omega
)})]/4 $, where $\Omega =\theta $, $\phi (\alpha ,t)$. For $k\ll 1$, this is
approximately $(\sigma _{q}^{2}|\xi (t)|^{2}/\sigma ^{2}+\Omega ^{2})/4$.
Multiplying $(\sigma _{q}^{2}|\xi (t)|^{2}/\sigma ^{2}+\Omega ^{2})/4$ and $%
\kappa \exp (-\kappa t)$ results in the photon arrival rate density $D(t)=$ $%
\kappa \exp (-\kappa t)(\sigma _{q}^{2}|\xi (t)|^{2}/\sigma ^{2}+\Omega
^{2})/(4P)$ in OC, where
\begin{equation}
P=(1/4)\int_{0}^{\infty }\kappa \exp (-\kappa t)(\sigma _{q}^{2}|\xi
(t)|^{2}/\sigma ^{2}+\Omega ^{2})dt  \label{ddd}
\end{equation}%
is the overall probability of a single photon successfully generating the
superposition state of $|n\rangle $ and $(c+c^{\dagger })|n\rangle $. Figure
4(b) and Figure 5(b) show the photon arrival rate density $D(t)$ for the
first and the second proposed schemes, respectively. They show that in the
bad-cavity limit $\kappa >\omega _{m}$, i.e., non-sideband resolved regime,
as the decay rate $\kappa $ of the cavity increases, $D(t)$ become
increasingly concentrated at time near $t=0$.
\begin{figure}[tp]
\includegraphics[width=1.65in]{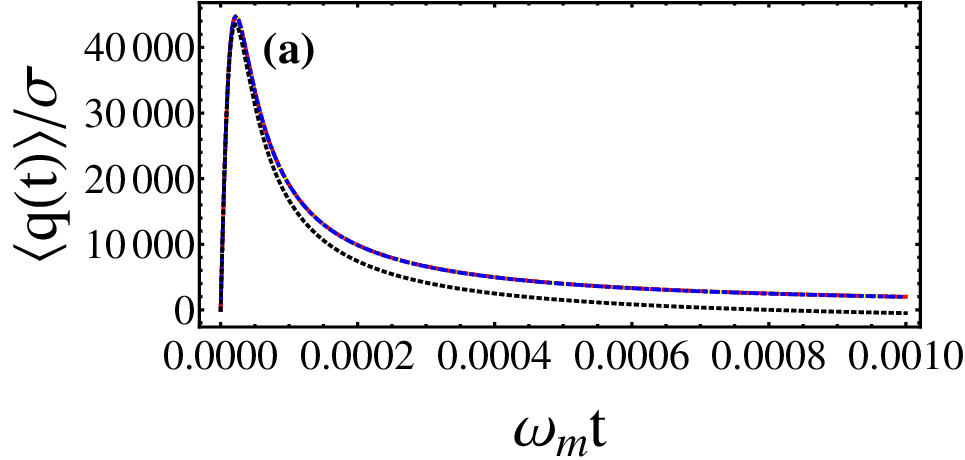} %
\includegraphics[width=1.65in]{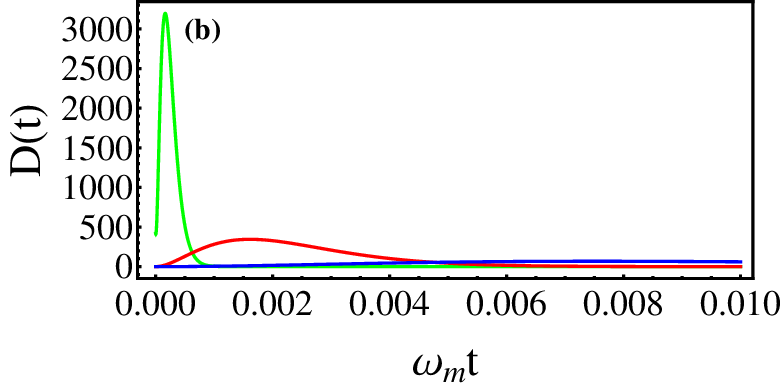}
\caption{(a) Average displacement $\langle q(t)\rangle /\protect\sigma $ at
time $t\ll 1$ with $k=0.005$, $\protect\theta =0.005$ and $\protect\omega %
_{m}=9\protect\pi $ kHz (room temperature 300K) for different $\protect%
\gamma =0$ (yellow line), $0.005$ (red line), $50$ (blue line) and $5\times
10^{3}$ (black line). (b) Photon arrival probability density $D(t)$ vs
arrival time for $\protect\theta $ ($\protect\theta =k$) with $\protect%
\kappa =1.2\times 10^{2}\protect\omega _{m}$ (blue line), $1.2\times 10^{3}%
\protect\omega _{m}$ (red line) and $1.2\times 10^{4}\protect\omega _{m}$
(green line).}
\end{figure}

For a repeated experimental set up with identical conditions, the "average"
displacement of the pointer is given by
\begin{equation}
\overline{\langle q(t)\rangle }=\int_{0}^{\infty }D(t)\langle q(t)\rangle dt,
\label{eee}
\end{equation}%
where $\langle q(t)\rangle $ is the same as $\langle q(t)\rangle $ in Eq. (%
\ref{gg}). At room temperature $T=300K$, we use a mechanical resonator with
mechanical frequency $f_{m}=4.5$ kHz and effective mass $m=100$ ng \cite%
{Pepper12}, indicating that $z=0.999999999$, $\sigma =4.32$ fm (femtometer).
So the maximal amplification value $\sigma _{q}=0.26$ nm. If $T=1500K$, $%
\sigma _{q}=0.5$ nm \cite{Temperature} For the first scheme, with $\kappa $ $%
=1.2\times 10^{4}\omega _{m}$, $\overline{\langle q(t)\rangle }=11577\sigma $
if $k=0.005$, $\theta =$ $0.005$, and for the second scheme, with $\kappa $ $%
=2\times 10^{4}\omega _{m}$, $\overline{\langle q(t)\rangle }=44704\sigma $
if $k=0.005$, $|\alpha |=\sigma _{q}/(2\sigma )$, $\beta =0$. Now we compare
these amplification results with the maximal unamplified value $4k\sigma
=86.4$ am (attometer) caused by the radiation pressure of single photon in
cavity A (amplification without the postselection, see Appendix F),
therefore the amplification factor is $Q=\overline{\langle q(t)\rangle }%
/(4k\sigma )$ which are $578850$ for the first scheme and $2235200$ for the
second scheme.

We then give the experimental requirements for the optomechanical device at
room temperature $T=300K$. According to Eq. (\ref{ddd}), $P$ that we need is
common, though the precise value of which depend on the dark count rate of
the detector and the stability of the setup. At room temperature $T=300K$,
for the first scheme, $P$ is approximately $6.94k^{2}$ (see Appendix H) for
a device with $\kappa =1.2\times 10^{4}\omega _{m}$ when $\theta =0.005$.
The window that detectors need to open for photons is approximately $%
1/\kappa $, requiring the dark count rate being lower than $6.94k^{2}\kappa $%
. The dark count rate of the best silicon avalanche photodiode is about $%
\sim 2$ Hz. So we require $k\geq 0$ for a $4.5$ kHz device, i.e. proposed
device no. 2 from \cite{Pepper12}, but with optical finesse $F$ reduced to $%
2800$ and cavity length being $0.5$ mm. For the second scheme, $P$ is
approximately $5k^{2}$ (see Appendix H) for a device with $\kappa =2\times
10^{4}\omega _{m}$ when $|\alpha |=\sigma _{q}/(2\sigma )$, $\beta =0$.
Because the dark count rate $2 $ Hz of the detector is lower than $%
5k^{2}\kappa $, we require $k\geq 0.000026$ for the same $4.5$ kHz device,
but with optical finesse $F$ reduced to $3000$ and cavity length being $0.3$
mm. Therefore, the implementation of the schemes provided here are feasible
to observe the impact of a single photon on a room-temperature mechanical
oscillator in experiment.
\begin{figure}[tp]
\includegraphics[width=1.65in]{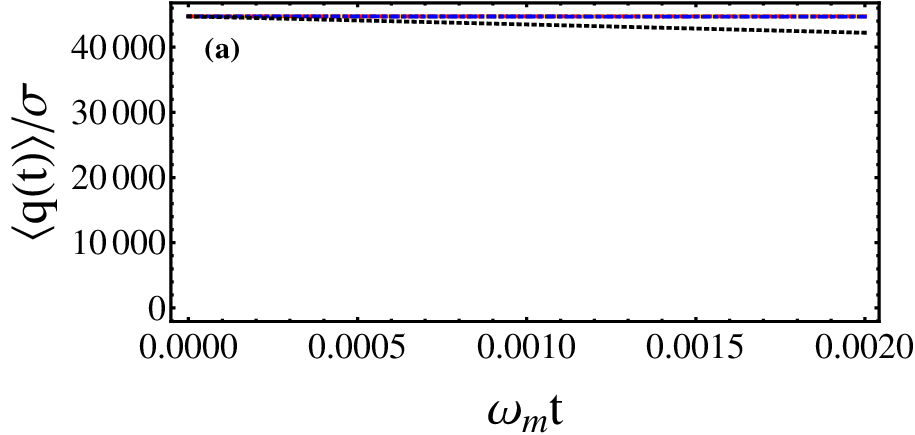} %
\includegraphics[width=1.65in]{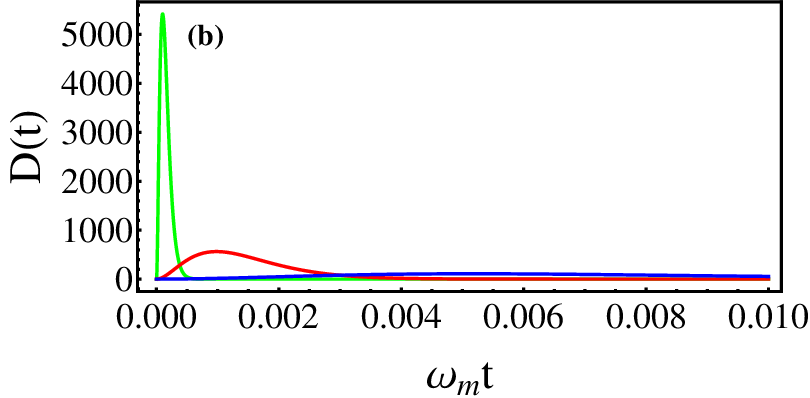}
\caption{(a) Average displacement $\langle q(t)\rangle /\protect\sigma $ at
time $t\ll 1$ with $k=0.005$, $|\protect\alpha |=(\frac{1+z}{1-z})^{1/2}/2$,
$\protect\beta =0$ and $\protect\omega _{m}=9\protect\pi $ kHz (room
temperature 300K) for different $\protect\gamma =0$ (yellow line), $0.005$
(red line), $0.5$ (blue line) and $50$ (black line). (b) Photon arrival
probability density $D(t)$ vs arrival time for $|\protect\alpha |=(\frac{1+z%
}{1-z})^{1/2}/2$, $\protect\beta =0$ with $\protect\kappa =2\times 10^{2}%
\protect\omega _{m}$ (blue line), $2\times 10^{3}\protect\omega _{m}$ (red
line) and $2\times 10^{4}\protect\omega _{m}$ (green line).}
\end{figure}

\section{Conclusion}

In this paper, we considered using thermal state to enhance the
amplification limit of the mechanical oscillator's displacement after the
postselection, and the maximal amplification value can reach the level of
thermal fluctuation, indicating that constantly improving the amplification
limit with the temperature increasing. In other words, thermal noise effect
of the pointer is beneficial for weak measurement amplification. The
mechanism behind the amplification is attributed to the superposition
between the number state $|n\rangle $ and the state $(c+c^{\dagger
})|n\rangle $ (unnormalized) of the postselection pointer and the classical
statistical properties of thermal state itself. To this end, we proposed two
different schemes for experimental implementations with optomechanical
system, and show that the amplification that occurs at time near $\omega
_{m}t=0$ is important for bad cavities with non-sideband resolved regime,
which means that our proposed two schemes are feasible to observe the impact
of a single photon on a room-temperature mechanical oscillator under current
experimental condition. Moreover, we have provided enough theoretical
toolbox \cite{Li=15,Li-15} to amplify the weaker effect in one-photon
weak-coupling optomechanics, which may be employed to explore the faint
gravitational effect.

\section{ACKNOWLEDGMENT}

This work was supported by the Natural Science Foundation of Shaanxi
Province (Grant No. 2018JQ1056), the Doctoral Scientific Research Foundation
of Yan'an University (Grant No. YDBK2016-04) and Youth Foundation of Yan'an
University (Grant No. YDQ2017-09). \newline

\appendix

\section{Amplification displacement of postselected weak measurement with
any state pointer}

The interaction Hamiltonian between the system and the pointer is
\begin{equation}
H_{int}=\chi (t)A\otimes q.  \label{eq:1-1}
\end{equation}%
Suppose the initial state of the system is $|\Phi _{i}\rangle
=(|a_{1}\rangle +|a_{2}\rangle )/\sqrt{2}$, and the initial state of the
pointer is $\rho _{m}$. The system is postselected in the state $|\Phi
_{p}\rangle =\cos \theta _{p}|a_{1}\rangle -e^{i\varphi }\sin \theta
_{p}|a_{2}\rangle $ after the interaction (\ref{eq:1-1}), and the pointer
collapses to the state (unnormalized)
\begin{align}
\rho _{pm}& =\langle \Phi _{p}|\exp (-i\chi Aq)|\Phi _{i}\rangle \langle
\Phi _{i}|\rho _{m}\exp (i\chi Aq)|\Phi _{p}\rangle  \notag \\
& =[\cos \theta _{p}\exp (-i\chi a_{1}q)-e^{-i\varphi }\sin \theta _{p}
\notag \\
& \exp (-i\chi a_{2}q)]\rho _{m}[\cos \theta _{p}\exp (i\chi a_{1}q)  \notag
\\
& -e^{i\varphi }\sin \theta _{p}\exp (i\chi a_{2}q)]/2,  \label{eq:1-2}
\end{align}%
where $\hat{q}=$ $\sigma (\hat{c}+\hat{c}^{\dagger })$, $\sigma $ is the
zero-point fluctuation. The success postselection probability is $%
P_{s}=Tr(\rho _{pm})$. However, if $\chi $ $\ll 1$ and $\varphi \ll 1$, and
when $\theta _{p}=\pi /4-\varepsilon $ with $\varepsilon \ll 1$, $\rho _{pm}$
(\ref{eq:1-2}) is approximately
\begin{eqnarray}
\rho _{pm} &\approx &(1/4)[2\varepsilon +i\varphi +i\chi (a_{2}-a_{1})q]\rho
_{m}[2\varepsilon  \notag \\
&-&i\varphi -i\chi (a_{2}-a_{1})q].  \label{eq:1-4}
\end{eqnarray}

The average displacement of the pointer observable $M$ ($M=p,q$) is
\begin{equation}
\langle M\rangle =Tr(M\rho _{pm})/Tr(\rho _{pm})-Tr(M\rho _{m}).
\label{eq:1-5}
\end{equation}%
Note that
\begin{align}
Tr(M\rho _{pm})& \approx \lbrack (4\varepsilon ^{2}+\varphi ^{2})\langle
M\rangle _{\rho _{m}}+i2\varepsilon \chi (a_{2}-a_{1})  \notag \\
& \langle \lbrack M,q]\rangle _{\rho _{m}}+\varphi \chi (a_{2}-a_{1})\langle
\{M,q\}\rangle _{\rho _{m}}  \notag \\
& +\chi ^{2}(a_{2}-a_{1})^{2}\langle qMq\rangle _{\rho _{m}}]/4,
\label{eq:1-6}
\end{align}%
and the normalized coefficient is

\begin{align}
A_{0}& =Tr(\rho _{pm})\approx \lbrack (4\varepsilon ^{2}+\varphi
^{2})+2\varphi \chi (a_{2}  \notag \\
& -a_{1})\langle q\rangle _{\rho _{m}}+\chi ^{2}(a_{2}-a_{1})^{2}\langle
q^{2}\rangle _{\rho _{m}}]/4,  \label{eq:1-7}
\end{align}%
where $Tr(\cdot \rho _{m})$ as $\langle \cdot \rangle _{\rho _{m}}$ for
short throughout the paper.

By substituting (\ref{eq:1-6}) and (\ref{eq:1-7}) into (\ref{eq:1-5}), we
find that
\begin{align}
\langle M\rangle & =A_{0}^{-1}[(4\varepsilon ^{2}+\varphi ^{2})\langle
M\rangle _{\rho _{m}}+i2\varepsilon \chi (a_{2}-a_{1})  \notag \\
& \langle \lbrack M,q]\rangle _{\rho _{m}}+\varphi \chi (a_{2}-a_{1})\langle
\{M,q\}\rangle _{\rho _{m}}  \notag \\
& +\chi ^{2}(a_{2}-a_{1})^{2}\langle qMq\rangle _{\rho _{m}}]/4-\langle
M\rangle _{\rho _{m}},  \label{eq:1-8}
\end{align}%
where $[\cdot ]$ and $\{\cdot \}$ denote commutation and anticommutation
rules, respectively. (\ref{eq:1-8}) is the average displacement of the any
pointer. If the initial state of the pointer $\rho _{m}$ satisfy the
symmetry condition, i.e, $F(-x)=F(x)$, the expression (\ref{eq:1-8}) becomes

\begin{eqnarray}
\langle M\rangle &=&A^{-1}(i2\varepsilon \chi (a_{2}-a_{1})\langle \lbrack
M,q]\rangle _{\rho _{m}}  \notag \\
&+&\varphi \chi (a_{2}-a_{1})\langle \{M,q\}\rangle _{\rho _{m}})/4,
\label{eq:1-9}
\end{eqnarray}%
where $A=[4\varepsilon ^{2}+\varphi ^{2}+\chi ^{2}(a_{2}-a_{1})^{2}\langle
q^{2}\rangle _{\rho _{m}}]/4$ is a normalized coefficient. It is obvious
that the displacement is determined by $i2\varepsilon \chi
(a_{2}-a_{1})\langle \lbrack M,q]\rangle _{\rho _{m}}$ and $\varphi \chi
(a_{2}-a_{1})\langle \{M,q\}\rangle _{\rho _{m}}$. The former and latter are
both caused by interference term of this state (\ref{eq:1-4}). In other
words, the key to understand the amplification is the coherence
(superposition) between the different states in the pointer after the
postselection.

There are two cases for Eq. (\ref{eq:1-9}): one is that when $\varphi =0$
and $\varepsilon \neq 0$, (\ref{eq:1-9}) becomes

\begin{equation}
\langle M\rangle =A_{1}^{-1}i2\varepsilon \chi (a_{2}-a_{1})\langle \lbrack
M,q]\rangle _{\rho _{m}},  \label{eq:1-10}
\end{equation}%
where $A_{1}=[4\varepsilon ^{2}+\chi ^{2}(a_{2}-a_{1})^{2}\langle
q^{2}\rangle _{\rho _{m}}]/4$ is a normalized coefficient. (\ref{eq:1-10})
correspond to the displacement space proportional to real weak value, the
result is holds up if and only if $M=p$; the other is that when $\varphi
\neq 0$ and $\varepsilon =0$, (\ref{eq:1-9}) becomes

\begin{equation}
\langle M\rangle =A_{2}^{-1}\varphi \chi (a_{2}-a_{1})\langle \{M,q\}\rangle
_{\rho _{m}},  \label{eq:1-11}
\end{equation}%
where $A_{2}=[\varphi ^{2}+\chi ^{2}(a_{2}-a_{1})^{2}\langle q^{2}\rangle
_{\rho _{m}}]/4$ is a normalized coefficient. (\ref{eq:1-11}) correspond to
the displacement space proportional to imaginary weak value, the result is
holds up if and only if $M=q.$

\subsection*{Amplification displacement based on a thermal pointer}

If we consider $\rho _{m}$ is a thermal state $\rho _{th}(z)$ (\ref{II}) in
the mian text, the final total pointer state after the postselection is
(normalized)
\begin{equation}
\rho _{pm}=B_{tot}^{-1}(1-z)\sum_{n=0}z^{n}|\psi _{m}(n)\rangle _{\eta \ll
1}\langle \psi _{m}(n)|_{\eta \ll 1},  \label{eq:1-12}
\end{equation}%
where $B_{tot}=(\sigma ^{2}\varphi ^{2}+4\sigma ^{2}\varepsilon ^{2}+\sigma
_{q}^{2}(a_{2}-a_{1})^{2}\eta ^{2})/4\sigma ^{2}$ is a normalized
coefficient for $\rho _{pm}$, and $\sigma _{q}=\coth ^{1/2}(\beta \omega
_{m}/2)\sigma $ represents thermal fluctuations of the position $q$ space.

Substituting $|\psi _{m}(n)\rangle _{\eta \ll 1}$ into Eq. (\ref{eq:1-5})
and $M=q$, when $\varphi \neq 0$ and $\varepsilon =0$, we obtain the
displacement of the pointer for $\psi _{m}(n)\rangle _{\eta \ll 1}$
\begin{eqnarray}
\langle q\rangle _{n} &=&Tr(q|\psi _{m}(n)\rangle _{\eta \ll 1}\langle \psi
_{m}(n)|_{\eta \ll 1})  \notag \\
&=&B_{n}^{2}CTr(\{M,q\}|n\rangle _{m}\langle n|_{m})/\sigma  \notag \\
&=&\sigma B_{n}^{2}C(2n+1),  \label{eq:1-13}
\end{eqnarray}%
where $C=\varphi \eta (a_{2}-a_{1})/2$ and $B_{n}=2[\varphi ^{2}+\eta
^{2}(a_{2}-a_{1})^{2}(2n+1)]^{-1/2}$ is a normalization coefficient for $%
|\psi _{m}(n)\rangle _{\eta \ll 1}$. Therefore, the above formula is the
same as Eq. (\ref{ppk}) in main text.

For Eq. (\ref{eq:1-12}), the average displacement of the pointer in position
$q$ space will be \ \ \ \
\begin{eqnarray}
\langle q\rangle &=&\sum_{n=0}P_{n}\langle q\rangle _{n}  \notag \\
&=&C\sigma _{q}^{2}/(\sigma B_{tot}),  \label{eq:1-14}
\end{eqnarray}%
where $P_{n}=z^{n}(1-z)B_{n}^{-2}/B^{tot}$ is the classical statistical
probability for each state $|\psi _{m}(n)\rangle _{\eta \ll 1}$ in the
ensemble of the pure state $\left\{ P_{n},\psi _{m}(n)\rangle _{\eta \ll
1}\right\} $, and $\langle p\rangle =0$.

Special note is given here, suppose that $\rho _{m}=|0\rangle \langle 0|$
(ground state) or $|\alpha \rangle \langle \alpha |$ (coherent state), the
maximal amplification value of (\ref{eq:1-11}) is the ground state
fluctuation $\sigma $, which are exactly confirmed by Eq. (17) in Ref. \cite%
{Li14} and Eq. (25) in Ref. \cite{Li15}, respectively. When $\rho _{m}=S(\xi
)|\alpha \rangle \langle \alpha |S^{\dagger }(\xi )$, $S(\xi )=\exp (\xi
^{\ast }a^{2}/2-\xi a^{\dagger 2}/2)$ with $\xi =re^{i\theta }$, the maximal
amplification value of (\ref{eq:1-11}) is the squeezing ground-state
fluctuation $\pm e^{r}\sigma $, which is exactly confirmed by Eq. (15) in
Ref. \cite{Li-15}.

Substituting Eq. (\ref{ppi}) into Eq. (\ref{eq:1-5}) and $M=p$, when $%
\varphi =0$ and $\varepsilon \neq 0$, we obtain the average displacement of
the pointer in momentum $p$ space
\begin{equation}
\langle p\rangle =(a_{2}-a_{1})\varepsilon \eta /(2\sigma B_{tot}),
\label{lkk}
\end{equation}
which is the asymptotic solution and $\langle q\rangle =0$. From Eq. (\ref%
{lkk}), we can still get the maximal positive value $1/(2\sigma _{q})$ when $%
\varepsilon =\sigma _{q}(a_{2}-a_{1})\eta /(2\sigma )$ and the maximal
negative value $-1/(2\sigma _{q})$ when $\varepsilon =-\sigma
_{q}(a_{2}-a_{1})\eta /(2\sigma )$, respectively. Because $\coth ^{-1}(\beta
\omega _{m}/2)<1$, so $\left\vert \langle p\rangle \right\vert <1/(2\sigma )$
(zero-point fluctuation), implying that the maximal amplification of the
pointer's displacement in momentum space is less than zero-point
fluctuation, in sharp contrast to Eq. (\ref{eeee}) in the following Section
II which indicate that $\langle p\rangle =\pm 1/(2\sigma )$ when $%
\varepsilon =\pm (a_{2}-a_{1})\eta /2$. Therefore, thermal noise effect of
the pointer has a negative effect for the amplification of the displacement
proportional to reak weak value.

Although the displacement proportional to reak weak value has been amplified
using thermal state pointer, but it is far less than the larger uncertainty
(thermal fluctuation) of the pointer, indicating that mixed state pointer
with larger fluctuation is infeasible for the displacement proportional to
reak weak value. In other words, if mixed state pointer (e.g., thermal
state) didn't have any advantage over pure state pointer, it would be
pointless to study amplification with mixed state pointer.

\section{Weak value based on a thermal state pointer}

\bigskip According to the definition of weak value \cite{Aharonov88}
\begin{equation}
A_{w}=\frac{\langle \psi _{p}|A|\psi _{i}\rangle }{\langle \psi _{p}|\psi
_{i}\rangle },  \label{ppl}
\end{equation}%
where $|\psi _{i}\rangle $ and $|\psi _{p}\rangle $ is the preselected and
postselected state. In this case of using thermal state as a pointer, the
weak-value regime satisfies the condition $\eta (2n+1)^{1/2}\ll \varphi
,\varepsilon \ll 1$. When the postselection state of the system $|\psi
_{p}\rangle =(\cos (\pi /4-\varepsilon )|a_{1}\rangle _{s}-e^{i\varphi }\sin
(\pi /4-\varepsilon )|a_{2}\rangle _{s})$ is performed for the total system (%
\ref{III}):
\begin{align}
\rho _{pm}& =\langle \psi _{p}|\exp [-i\eta A(\hat{c}+\hat{c}^{\dagger
})]|\psi _{i}\rangle \langle \psi _{i}|\rho _{th}\exp [i\eta A(\hat{c}
\notag \\
& +\hat{c}^{\dagger })]|\psi _{p}\rangle  \notag \\
& \approx (1-z)\sum_{n=0}z^{n}[\langle \psi _{p}|\psi _{i}\rangle -i\eta
\langle \psi _{p}|A|\psi _{i}\rangle (\hat{c}  \notag \\
& +\hat{c}^{\dagger })]|n\rangle _{m}\langle n|_{m}[\langle \psi _{i}|\psi
_{p}\rangle +i\eta \langle \psi _{i}|A|\psi _{p}\rangle (\hat{c}+\hat{c}%
^{\dagger })]  \notag \\
& \approx (1-z)\sum_{n=0}z^{n}\langle \psi _{p}|\psi _{i}\rangle \langle
\psi _{i}|\psi _{p}\rangle \exp [-i\eta A_{w}(\hat{c}  \notag \\
& +\hat{c}^{\dagger })]|n\rangle _{m}\langle n|_{m}\exp [i\eta A_{w}^{\ast }(%
\hat{c}+\hat{c}^{\dagger })]  \label{be}
\end{align}%
with
\begin{equation}
A_{w}\approx \mathop{\rm Re}A_{w}+i\mathop{\rm Im}A_{w},  \label{kklj}
\end{equation}%
where $\mathop{\rm Re}A_{w}=2\varepsilon (a_{1}-a_{2})/(4\varepsilon
^{2}+\varphi ^{2})\ $and $\mathop{\rm Im}A_{w}=-\varphi
(a_{1}-a_{2})/(4\varepsilon ^{2}+\varphi ^{2})$.

Substituting Eq. (\ref{be}) into Eq. (\ref{eq:1-5}) and $\varepsilon =0$,
the the average displacement of the pointer in position $q$ space is
\begin{align}
\langle q\rangle & =\sigma (1-z)\sum_{n=0}z^{n}\langle n|_{m}\exp [i\eta
A_{w}^{\ast }(\hat{c}+\hat{c}^{\dagger })](c  \notag \\
& +c^{\dagger })\exp [-i\eta A_{w}(\hat{c}+\hat{c}^{\dagger })]|n\rangle
_{m}/[(1-z)  \notag \\
& \sum_{n=0}z^{n}\langle n|_{m}\exp [-i\eta (A_{w}-A_{w}^{\ast })(\hat{c}+%
\hat{c}^{\dagger })]|n\rangle _{m}].  \label{bbl}
\end{align}%
Changing to the $q$ representation in rectangular coordinate, this becomes
\begin{eqnarray}
\langle q\rangle &=&(1-z)\int_{-\infty }^{\infty }\sum_{n=0}z^{n}dq(q\phi
_{n}^{2}(q)\exp [-i\eta (A_{w}  \notag \\
&-&A_{w}^{\ast })q/\sigma ])/[(1-z)\int_{-\infty }^{\infty
}\sum_{n=0}z^{n}dq\phi _{n}^{2}(q)  \notag \\
&&\exp [-i\eta (A_{w}-A_{w}^{\ast })q/\sigma ]],  \label{bbj}
\end{eqnarray}%
and $\phi _{n}(q)$ is defined as%
\begin{equation}
\phi _{n}(q)=(2^{n}n!)^{-1/2}H_{n}(q/\sqrt{2}\sigma )\phi _{0}(q),
\label{bbh}
\end{equation}%
where $\phi _{0}(q)=(2\pi \sigma ^{2})^{-1/4}$ $\exp [-q^{2}/(4\sigma
^{2})]) $ and $H_{n}$ is Hermite Polynomial.

Using Mehler's Hermite Polynomial Formula \cite{Gradshteyn65}
\begin{eqnarray}
&\sum\limits_{n=0}^{\infty
}&H_{n}(x)H_{n}(y)w^{n}(2^{n}n!)^{-1}=(1-w^{2})^{-1/2}  \notag \\
&&\exp [(2xyw-(x^{2}+y^{2})w^{2})/(1-w^{2})],  \label{bby}
\end{eqnarray}%
and
\begin{eqnarray}
&&\int_{-\infty }^{\infty }dx(x\exp [-x^{2}]\exp [mx])  \notag \\
&=&\frac{d}{dm}\int_{-\infty }^{\infty }dx(\exp [-x^{2}]\exp [mx])],
\label{bbt}
\end{eqnarray}%
then Eq. (\ref{bbj}) becomes%
\begin{equation}
\langle q\rangle =2\chi \mathop{\rm Im}A_{w}\sigma _{q}^{2}.  \label{bj}
\end{equation}%
From Eq. (\ref{bj}), it can be seen that $\langle q\rangle $ is proportional
to the square of thermal fluctuation and is imaginary in position $q$ space,
which is the generalization of the result of Eq. (10) in \cite{Josza07}.
Therefore, thermal noise effect of the pointer is beneficial for weak
measurement amplification. But $\langle q\rangle $ for the weak-value
amplification is not the optimal displacement, i.e., not the maximal
amplification value. The maximal amplification value is $\pm \sigma _{q}$
(thermal fluctuation) in the main text.

Substituting Eq. (\ref{be}) into Eq. (\ref{eq:1-5}) and $\varphi =0$, the
average displacement of the pointer in momentum $p$ space is
\begin{align}
\langle p\rangle & =-i(2\sigma )^{-1}(1-z)\sum_{n=0}z^{n}\langle n|_{m}\exp
[i\eta A_{w}^{\ast }(\hat{c}  \notag \\
& +\hat{c}^{\dagger })](c-c^{\dagger })\exp [-i\eta A_{w}(\hat{c}+\hat{c}%
^{\dagger })]|n\rangle _{m}  \notag \\
& =-\chi \mathop{\rm Re}A_{w},  \label{bhj}
\end{align}%
which is exactly the same weak values as a pure Gaussian pointer state \cite%
{Aharonov88}.

\section{Fock state veiw of the standard weak measurement with a ground
state pointer}

We consider the Hamiltonian (\ref{I}) in the main text. If the initial state
of the system is $|\psi _{i}\rangle =(|a_{1}\rangle +|a_{2}\rangle )/\sqrt{2}
$, where $|a_{1}\rangle $ and $|a_{2}\rangle $ is eigenstates of $A$. Any
Gaussian can be seen as the ground state of a fictional harmonic oscillator
Hamiltonian \cite{Scully1997}. Suppose the initial pointer state is the
ground state $|0\rangle _{m}$. Then weakly couples them using the
interaction Hamiltonian (\ref{eq:1-1}), the time evolution of the total
system is given by
\begin{align}
& U(t)|\psi _{i}\rangle |0\rangle _{m}  \notag \\
& =\exp [-i\eta A(\hat{c}+\hat{c}^{\dagger })]|\psi _{i}\rangle |0\rangle
_{m}  \notag \\
& =[|a_{1}\rangle D(-ia_{1}\eta )+|a_{2}\rangle D(-ia_{2}\eta ]|0\rangle
_{m})/\sqrt{2},  \label{aaaa}
\end{align}%
where $U(t)=e^{-i\chi \hat{A}\hat{q}}$.

When the postselection $|\psi _{p}\rangle =[\cos (\pi /4-\varepsilon
)|a_{1}\rangle -e^{i\varphi }\sin (\pi /4-\varepsilon )|a_{2}\rangle ]$ with
$\varepsilon \ll 1$ is performed for the total system (\ref{aaaa}), i.e., $%
\langle \psi _{p}|\psi _{i}\rangle \approx \varepsilon +i\varphi /2$, then
the final state of the pointer is
\begin{eqnarray}
&&(1/\sqrt{2})[\cos (\pi /4-\varepsilon )D(-ia_{1}\eta )-e^{-i\varphi }\sin
(\pi /4  \notag \\
&&-\varepsilon )D(-ia_{2}\eta )]|0\rangle _{m}.  \label{bbbb}
\end{eqnarray}

For Eq. (\ref{bbbb}), when $\varphi \ll 1$, $\varepsilon \ll 1$ and $\eta
\ll 1$, we can then perform a small quantity expansion about $\eta $ and $%
\varepsilon $ till the second order, and then obtain
\begin{equation}
\lbrack (2\varepsilon +i\varphi )|0\rangle _{m}+i\eta (a_{2}-a_{1})|1\rangle
_{m}]/2.  \label{cccc}
\end{equation}

Substituting Eq. (\ref{cccc}) into Eq. (\ref{eq:1-5}), in this case of the
near-orthogonal postselection, i.e., $\langle \psi _{p}|\psi _{i}\rangle
\approx \varepsilon $ (real), we can find that
\begin{equation}
\langle \hat{p}\rangle =2(a_{2}-a_{1})\varepsilon \eta /(4\varepsilon
^{2}\sigma +(a_{2}-a_{1})^{2}\eta ^{2}\sigma ).  \label{eeee}
\end{equation}%
and $\langle \hat{q}\rangle =0$.

When $2\varepsilon =\pm (a_{2}-a_{1})\eta $, we will have the largest
displacement $\pm 1/(2\sigma )$ in momentum $p$ space and when $\varepsilon
=0$, indicating that the postselected state of the system is orthogonal to
the initial state of the system, i.e., $\langle \psi _{p}|\psi _{i}\rangle
=0 $, the displacement of the pointer in momentum $p$ space is $0$. This
amplification result is due to the superposition of $|0\rangle _{m}$ and $%
|1\rangle _{m}$. However, the displecement of the pointer in position $q$
space is always $0$.

Substituting Eq. (\ref{cccc}) into Eq. (\ref{eq:1-5}), in this case of the
near-orthogonal postselection, i.e., $\langle \psi _{p}|\psi _{i}\rangle
\approx i\varphi /2$ (imaginary), we obtain
\begin{equation}
\langle q\rangle =2\sigma (a_{2}-a_{1})\varphi \eta /[\varphi
^{2}+(a_{2}-a_{1})^{2}\eta ^{2}]  \label{eefff}
\end{equation}%
and $\langle p\rangle =0$.

When $\varphi =\pm (a_{2}-a_{1})\eta $ we will have the largest displacement
$\pm \sigma $ in position $q$ space and when $\varphi =0$, indicating that
the postselected state of the system is orthogonal to the initial state of
the system, i.e., $\langle \psi _{p}|\psi _{i}\rangle =0$, the displacement
of the pointer in position $q$ space is $0$. This amplification result is
due to the superposition of $|0\rangle _{m}$ and $|1\rangle _{m}$. However,
the displecement of the pointer in momentum $p$ space is always $0$.

Obviously, the mechanism behind the amplification with Gaussian pointer \cite%
{Aharonov88} is also regarded as the superposition of $|0\rangle $ and $%
|1\rangle $ of the pointer in fock space. Therefore, the standard scenario
of weak measurement \cite{Aharonov88} can be also shown and understood by
the Fock-state view where the initial state of the pointer is a ground state
\cite{Simon11}. It give a view of the relationship between the weak
measurement and other measurement techniques.

\section{Amplification using a phase shifter $\protect\theta $ in
optomechanics}

\bigskip According to the results of Ref. \cite{Bose97,Mancini97}, the time
evolution operator of the Hamiltonian (\ref{ee}) in the main text is given
by
\begin{eqnarray}
U(t) &=\exp [-ir(a^{\dag }a+b^{\dag }b)\omega _{m}t]\exp [i(a^{\dag
}a)^{2}\phi (t)]  \notag \\
&\exp [a^{\dag }a(\xi (t)c^{\dag }-{\xi }^{\ast }(t)c)]\exp [-ic^{\dag
}c\omega _{m}t],  \label{l}
\end{eqnarray}
where $\phi (t)=k^{2}(\omega _{m}t-\sin \omega _{m}t)$, $\xi
(t)=k(1-e^{-i\omega _{m}t})$, $r=\omega _{0}/\omega _{m}$, $k=g/\omega _{m}$
is the scaled coupling parameter.

Suppose that one photon is input into the interferometer, and after the
first beam splitter and a phase shifter $\theta $ the initial state of the
photon is $|\psi _{i}(\theta )\rangle =(1/\sqrt{2})(e^{i\theta }|1\rangle
_{A}|0\rangle _{B}+|0\rangle _{A}|1\rangle _{B})$. The mirror is initialised
in thermal state $\rho _{th}(z)$. After weakly coupled interacting (\ref{l})
between one photon and the mirror, the time evolution of the total system
leads to a state given by

\begin{eqnarray}
\rho (z) &=&(1-z)\sum_{n=0}z^{n}[|1\rangle _{A}|0\rangle _{B}e^{i(\phi
(t)+\theta )}D(\xi )  \notag \\
&+&|0\rangle _{A}|1\rangle _{B}]|n\rangle _{m}\langle n|_{m}[\langle
1|_{A}\langle 0|_{B}e^{-i(\phi (t)+\theta )}  \notag \\
&+&D^{\dagger }(\xi )\langle 0|_{A}\langle 1|_{B}]/2.  \label{k}
\end{eqnarray}%
When a photon is detected in the dark port, in the language of weak
measurement the postselected state of one photon is $|\psi _{p}\rangle
=(|1\rangle _{A}|0\rangle _{B}-|0\rangle _{A}|1\rangle _{B})/\sqrt{2}$,
which is nonorthogonal to $|\psi _{i}(\theta )\rangle $, i.e., $\langle \psi
_{f}|\psi _{i}(\theta )\rangle \approx i\theta /2$. Then the reduced state
of the mirror after the postselection for each $n$ component of the pointer
state is
\begin{align}
|\psi _{1}(n)\rangle & =\langle \psi _{p}|(|1\rangle _{A}|0\rangle
_{B}e^{i(\phi (t)+\theta )}D(\xi )  \notag \\
& +|0\rangle _{A}|1\rangle _{B})|n\rangle _{m}/\sqrt{2}  \notag \\
& =[e^{i(\phi (t)+\theta )}D(\xi )|n\rangle _{m}-|n\rangle _{m}]/2.
\label{q}
\end{align}

Therefore, this is Eq. (\ref{dd}) in the main text.

For Eq. (\ref{q}), over all $n$ component, then the final total state of the
pointer is $\rho _{m}^{pha}=(1-z)\sum_{n=0}z^{n}|\psi _{1}(n)\rangle \langle
\psi _{1}(n)|.$ Substituting Eq. (\ref{q}) into Eq. (\ref{eq:1-5}), we can
follow a two-step procedure to obtain the average displacement of the
mirror: first, calculate the numerator of equation (\ref{eq:1-5}), then
calculate the denominator of equation (\ref{eq:1-5}).

For the numerator of Eq. (\ref{eq:1-5}), we obtain%
\begin{align}
& \langle n|D^{\dagger }(\xi )qD(\xi )|n\rangle =\sigma \langle n|D^{\dagger
}(\xi )(c+c^{\dagger })D(\xi )|n\rangle  \notag \\
& =\sigma (\xi (t)+\xi ^{\dagger }(t)),  \label{ggaa}
\end{align}%
using $D^{\dagger }(\alpha )cD(\alpha )=c+\alpha $, $D^{\dagger }(\alpha
)c^{+}D(\alpha )=c^{\dagger }+\alpha ^{\ast }$, and
\begin{equation}
\langle n|q|n\rangle =\sigma \langle n|(c+c^{\dagger })|n\rangle =0,
\label{ggbb}
\end{equation}

\begin{align}
&e^{i(\phi (t)+\theta )}\langle n|qD(\xi )|n\rangle  \notag \\
&=\sigma e^{i(\phi(t)+\theta )}\langle n|(c+c^{\dagger })D(\xi )|n\rangle ,
\label{ggcc}
\end{align}

\begin{align}
&e^{-i(\phi(t)+\theta)}\langle n|D^{\dagger}(\xi)q|n\rangle  \notag \\
&=\sigma e^{-i(\phi(t)+\theta)}\langle
n|D^{\dagger}(\xi)(c+c^{\dagger})|n\rangle ,  \label{ggdd}
\end{align}

For Eq. (\ref{ggcc}), and using
\begin{align}
& \langle l|D(\alpha )|n\rangle =\sqrt{n!/l!}\alpha ^{(l-n)}\exp (-|\alpha
|^{2}/2)\times  \notag \\
& L_{n}^{(l-n)}(|\alpha |^{2}),(l\geq n),  \label{qq}
\end{align}%
and
\begin{align}
& \langle l|D^{\dagger }(\alpha )|n\rangle =\sqrt{n!/l!}(-\alpha
)^{(l-n)}\exp (-|-\alpha |^{2}/2)\times  \notag \\
& L_{n}^{(l-n)}(|-\alpha |^{2}),(l\geq n),  \label{pp}
\end{align}%
where $L_{n}^{k}(x)$ is an associated Laguerre polynomial \cite{Gradshteyn65}%
, we find that
\begin{align}
& e^{i(\phi (t)+\theta )}\langle n|qD(\xi )|n\rangle =\sigma e^{i(\phi
(t)+\theta )}[(n+1)^{1/2}\langle n+1|  \notag \\
& D(\xi )|n\rangle +n^{1/2}\langle n-1|D(\xi )|n\rangle ]  \notag \\
& =\sigma e^{i(\phi (t)+\theta )}D_{n+1,n}+\sigma e^{i(\phi (t)+\theta
)}D_{n,n-1}^{\dagger \ast }  \label{qqq}
\end{align}%
with
\begin{equation}
D_{n+1,n}=\xi \exp (-|\xi |^{2}/2)L_{n}^{1}(|\xi |^{2}),n\geq 0  \label{ppoo}
\end{equation}%
and
\begin{align}
& D_{n,n-1}^{\dagger }=-\xi \exp (-|-\xi |^{2}/2)L_{n}^{1}(|-\xi |^{2}),
\notag \\
& n\geq 1.  \label{ppio}
\end{align}%
Using identity
\begin{equation}
\sum_{n=0}^{\infty }L_{n}^{k}(x)z^{n}=(1-z)^{-k-1}\exp (-xz/(1-z)),
\label{pppq}
\end{equation}%
we have the following result
\begin{align}
& (1-z)\sum_{n=0}^{\infty }z^{n}D_{n+1,n}  \notag \\
& =\xi \exp [-\sigma _{q}^{2}|\xi |^{2}/(2\sigma ^{2})]/(1-z)  \label{ppol}
\end{align}%
Seting $n=n^{^{\prime }}+1$ and using Eq. (\ref{pppq}),
\begin{align}
& (1-z)\sum_{n=0}^{\infty }z^{n}D_{n,n-1}^{\dagger \ast }  \notag \\
& =(1-z)\sum_{n^{^{\prime }}=0}^{\infty }z^{n^{^{\prime }}+1}D_{n^{^{\prime
}}+1,n^{^{\prime }}}^{\dagger \ast }  \notag \\
& =-z\xi ^{\ast }\exp [-\sigma _{q}^{2}|\xi |^{2}/(2\sigma ^{2})]/(1-z)
\label{qqqn}
\end{align}%
Then we have
\begin{align}
& (1-z)e^{i(\phi (t)+\theta )}\sum_{n=0}^{\infty }\langle n|qD(\xi
)|n\rangle =\sigma \lbrack \xi (t)  \notag \\
& \exp (-\sigma _{q}^{2}|\xi |^{2}/(2\sigma ^{2})+i\phi (t)+i\theta )/(1-z)-z
\notag \\
& \xi ^{\ast }\exp (-\sigma _{q}^{2}|\xi |^{2}/(2\sigma ^{2})+i\phi
(t)+i\theta )/(1-z)]  \label{ppr}
\end{align}

Next, for the denominator of Eq. (\ref{eq:1-5}), and using Eq. (\ref{qq})
and Eq. (\ref{pp}), we find that

\begin{align}
& e^{i(\phi (t)+\theta )}\langle n|D(\xi )|n\rangle  \notag \\
& =e^{i(\phi (t)+\theta )}\exp (-|\xi |^{2})L_{n}^{0}(|\xi |^{2}/2),n\geq 0
\label{ggccc}
\end{align}

and

\begin{align}
& e^{-i(\phi (t)+\theta )}\langle n|D^{\dagger }(\xi )|n\rangle  \notag \\
& =e^{-i(\phi (t)+\theta )}\exp (-|\xi |^{2})L_{n}^{0}(|\xi |^{2}/2),n\geq 0
\label{ggddd}
\end{align}

For Eq. (\ref{ggccc}), using identity (\ref{pppq}), we have the following
result%
\begin{align}
& (1-z)e^{i(\phi (t)+\theta )}\sum_{n=0}^{\infty }\langle n|D(\xi )|n\rangle
\notag \\
& =e^{i(\phi (t)+\theta )}\exp (-\sigma _{q}^{2}|\xi |^{2}/(2\sigma ^{2}))
\label{kkcc}
\end{align}

So we can obtain the average displacement of the mirror
\begin{align}
& \langle q(t)\rangle =(1-z)[\sum_{n=0}^{\infty }\langle n|D^{\dagger }(\xi
)qD(\xi )|n\rangle -e^{i(\phi (t)+\theta )}  \notag \\
& \sum_{n=0}^{\infty }\langle n|qD(\xi )|n\rangle -e^{-i(\phi (t)+\theta
)}\sum_{n=0}^{\infty }\langle n|D^{\dagger }(\xi )q|n\rangle ]/[2  \notag \\
& -(1-z)e^{i(\phi (t)+\theta )}\sum_{n=0}^{\infty }\langle n|D(\xi
)|n\rangle -(1-z)  \notag \\
& e^{-i(\phi (t)+\theta )}\sum_{n=0}^{\infty }\langle n|D^{+}(\xi )|n\rangle
]  \notag \\
& =\sigma \lbrack \xi +\xi ^{\ast }-(1-z)^{-1}[\Phi \xi +\Phi ^{\ast }\xi
^{\ast }  \notag \\
& -z(\Phi \xi ^{\ast }+\Phi ^{\ast }\xi )]]/(2-\Phi -\Phi ^{\ast }),
\label{r}
\end{align}%
where $\Phi =\exp (-\sigma _{q}^{2}|\xi |^{2}/(2\sigma ^{2})+i\phi
(t)+i\Omega )$ with $\Omega =\theta $. Therefore, this is Eq. (\ref{gg}) in
the main text. Note that the denominator $(2-\Phi -\Phi ^{\ast })/4$ is the
successful postselection probability being released from optomechanical
cavity after the time $t$.

\subsection*{Small quantity expansion about time for amplification}

However, in order to observe the amplification effects appearing at time
near $T=0$, for Eq. (\ref{q}) we can then perform a small quantity expansion
about time $T$ till the second order. Suppose that $|\omega _{m}t-T|\ll 1$,
i.e., $\omega _{m}t\ll 1$, $k\ll 1$ and $\theta \ll 1$, then we can obtain

\begin{align}
& \psi _{1}(n)\rangle _{\omega _{m}t\ll 1}\approx \lbrack (1+i\theta
)(1+ik\omega _{m}t(c+c^{\dagger }))|n\rangle  \notag \\
& -|n\rangle ]/2  \notag \\
& =B_{1}(n)[i\theta |n\rangle +ik\omega _{m}t(c+c^{\dagger }))|n\rangle ]/2,
\label{rr}
\end{align}%
where $B_{1}(n)=2[\theta ^{2}+k^{2}(\omega _{m}t)^{2}(2n+1)]^{-1/2}$ is a
normalization coefficient for each state $|\psi _{1}(n)\rangle _{\omega
_{m}t\ll 1}$. Therefore, this is Eq. (\ref{hh}) in the main text.

Substituting Eq. (\ref{rr}) into Eq. (\ref{eq:1-5}), then
\begin{equation}
\langle q(t)\rangle _{n}=\sigma B_{1}^{2}(n)\theta k\omega _{m}t(2n+1)/2.
\label{ii}
\end{equation}%
Therefore, this is the average displacement $\langle q(t)\rangle _{n}$ for $%
\psi _{1}(n)\rangle _{\omega _{m}t\ll 1}$ plotted in Figure 2(b) in main
text.

For Eq. (\ref{rr}), over all $n$ component, then the final total state of
the pointer is $\rho _{m}^{pha}=(1-z)\sum_{n=0}z^{n}\psi _{1}(n)\rangle
_{\omega _{m}t\ll 1}\langle \psi _{1}(n)|_{\omega _{m}t\ll 1}/B_{1}^{tot}$
and substituting it into Eq. (\ref{eq:1-5}), then
\begin{equation}
\langle q(t)\rangle _{\omega _{m}t\ll 1}=\theta k\omega _{m}t\sigma
_{q}^{2}/(\sigma B_{1}^{tot}),  \label{uu}
\end{equation}%
where $B_{1}^{tot}=[\theta ^{2}\sigma ^{2}+k^{2}(\omega _{m}t)^{2}\sigma
_{q}^{2}]/(4\sigma ^{2})$ is a normalized coefficient for $\rho _{m}^{pha}$.
Based on Eq. (\ref{uu}), we then obtain the maximal positive value $\sigma
_{q}$ (thermal fluctuation) or negative value $-\sigma _{q}$ when $\theta
=\pm k\omega _{m}t\sigma _{q}$/$\sigma $, respectively. Therefore, the $%
|\psi (n)\rangle $ components corresponding to the maximal positive and
negative amplification, respectively, are $|\psi _{1}(n)\rangle _{\max
,\omega _{m}t\ll 1}=[\sigma _{q}/\sigma \pm (c+c^{\dagger })]|n\rangle ]/%
\sqrt{2}$ (unnormalized). Then the mirror state achieving the maximal
positive and negative amplification, respectively, are $\rho
_{m}^{pha}=(1-z)\sum_{n=0}z^{n}|\psi _{1}(n)\rangle _{\max ,\omega _{m}t\ll
1}\langle \psi _{1}(n)|_{\max ,\omega _{m}t\ll 1}/(4B_{1}^{tot})$. It is
obvious that the amplification with thermal state pointer is much larger
than that with pure state pointer \cite{Aharonov88,Simon11,Li14,Li15} since
its maximal value is the ground state fluctuation $\sigma $. Therefore,
thermal noise effect of the pointer (mirror) is beneficial for the
amplification of the mirror's displacement.

\section{Amplification using a displaced thermal state in optomechanics}

Suppose that one photon is input into the interferometer, and after the
first beam splitter the initial state of the photon is $|\psi _{i}\rangle
=(|1\rangle _{A}|0\rangle _{B}+|0\rangle _{A}|1\rangle _{B})/\sqrt{2}$. The
mirror is initialised in displaced thermal state $\rho _{th}(z,\alpha )$.
When the photon interact weakly with the optomechanical system through (\ref%
{l}), the evolution state of the total system is given by
\begin{eqnarray}
\rho (z) &=&(1-z)\sum_{n=0}z^{n}[|1\rangle _{A}|0\rangle _{B}e^{i\phi
(t)}D(\xi )  \notag \\
&+&|0\rangle _{A}|1\rangle _{B}]D(\varphi )|n\rangle _{m}\langle
n|_{m}D^{\dagger }(\varphi )[e^{-i\phi (t)}  \notag \\
&&D^{\dag }(\xi )|1\rangle _{A}|0\rangle _{B}+\langle 0|_{A}\langle
1|_{B}]/2,  \label{v}
\end{eqnarray}%
where $\xi (t)=k(1-e^{-i\omega _{m}t})$ and $\phi (t)=k^{2}(\omega
_{m}t-\sin \omega _{m}t)$ with $k=g/\omega _{m}$.

When a photon is detected in the dark port, in the language of weak
measurement the postselected state of the one photon is $|\psi _{p}\rangle
=(|1\rangle _{A}|0\rangle _{B}-|0\rangle _{A}|1\rangle _{B})/\sqrt{2}$,
which is orthogonal to $|\psi _{i}\rangle $, i.e., $\langle \psi _{p}|\psi
_{i}\rangle =0$. Then the reduced state of the mirror after the
postselection for each $n$ component of the pointer state is given by

\begin{align}
&|\chi _{2}(n)\rangle=[\langle \psi _{p}|[|1\rangle _{A}|0\rangle
_{B}e^{i\phi (t)}D(\xi )D(\varphi )  \notag \\
&+|0\rangle _{A}|1\rangle _{B}D(\varphi )]|n\rangle _{m}]/\sqrt{2}  \notag \\
&=[e^{i\phi (t)}D(\xi )D(\varphi )|n\rangle _{m}-D(\varphi )|n\rangle
_{m}]/2.  \label{vv}
\end{align}

In order to make the analysis simple, we can displace the above state to the
origin point in phase space, defining $|\psi _{2}(n)\rangle =D^{\dag
}(\varphi )|\chi _{2}(n)\rangle $ and we can obtain

\begin{align}
& |\psi _{2}(n)\rangle =[e^{i\phi (t)}D^{\dagger }(\varphi )D(\xi )D(\varphi
)  \notag \\
& -D^{\dagger }(\varphi )D(\varphi )]|n\rangle _{m}/2  \notag \\
& =[e^{i(\phi (t)+\phi (\alpha ,t))}D(\xi )|n\rangle _{m}-|n\rangle _{m}]/2,
\label{vvv}
\end{align}%
where $\phi (\alpha ,t)=-i[\alpha \xi -\alpha ^{\ast }\xi ^{\ast }]$ is
obtained by using the property of the displacement operators $D(\alpha
)D(\beta )=\exp [\alpha \beta ^{\ast }-\alpha ^{\ast }\beta ]D(\beta
)D(\alpha )$, due to noncommutativity of quantum mechanics \cite{Li15}.

For Eq. (\ref{vvv}), over all $n$ component, then the final total state of
the pointer is
\begin{equation}
\rho _{m}^{dis}=(1-z)\sum_{n=0}z^{n}|\psi _{2}(n)\rangle \langle \psi
_{2}(n)|.  \label{vvvv}
\end{equation}
Therefore, this is Eq. (\ref{aaa}) in main text.

Substituting Eq. (\ref{vvvv}) into Eq. (\ref{eq:1-5}), then we show the
average displacement of the mirror's position

\begin{align}
& \langle q(t)\rangle =\sigma \lbrack \xi +\xi ^{\ast }-(1-z)^{-1}[\Phi \xi
\notag \\
& +\Phi ^{\ast }\xi ^{\ast }-z(\Phi \xi ^{\ast }+\Phi ^{\ast }\xi
)]]/(2-\Phi -\Phi ^{\ast }),  \label{rrvv}
\end{align}%
where $\Phi =\exp (-\sigma _{q}^{2}|\xi |^{2}/(2\sigma ^{2})+i\phi
(t)+i\Omega )$ with $\Omega =\phi (\alpha ,t)$. In order to obtain the above
result, here we use two equations,

\begin{align}
& \langle l|D(\alpha )|n\rangle =\sqrt{n!/l!}\alpha ^{l-n}\exp (-|\alpha
|^{2}/2)  \notag \\
& \times L_{n}^{(l-n)}(|\alpha |^{2}),(l\geq n)  \label{rrt}
\end{align}%
and
\begin{equation}
\sum_{n=0}^{\infty }L_{n}^{k}(x)z^{n}=(1-z)^{-k-1}\exp (-xz/(1-z)),
\label{rrtt}
\end{equation}%
where $L_{n}^{k}(x)$ is an associated Laguerre polynomial \cite{Gradshteyn65}%
. Note that the denominator of Eq. (\ref{rrvv}) $(2-\Phi -\Phi ^{\ast })/4$
is the successful postselection probability being released from
optomechanical cavity after the time $t$.

Therefore, Eq. (\ref{rrvv}) is the average displacement $\langle q(t)\rangle
$ of the mirror for the state $|\psi _{2}(n)\rangle $ plotted in Figure 3(a)
in main text.

\subsection*{Small quantity expansion about time for amplification}

However, in order to observe the amplification effects appearing at time
near $T=0$, for Eq. (\ref{vvv}) we can then perform a small quantity
expansion about time $T$ till the second order. Suppose that $|\omega
_{m}t-T|\ll 1$, i.e., $\omega _{m}t\ll 1$, $k\ll 1$ and $2k|\alpha |\zeta
\ll 1$, then we can obtain
\begin{align}
& \psi _{2}(n)\rangle _{\omega _{m}t\ll 1}\approx \lbrack (1+i2k|\alpha
|\zeta )(1+ik\omega _{m}t(c+c^{\dagger }))|n\rangle  \notag \\
& -|n\rangle ]/2  \notag \\
& =B_{2}(n)[i2k|\alpha |\zeta |n\rangle +ik\omega _{m}t(c+c^{\dagger
})|n\rangle ]/2,  \label{yynn}
\end{align}%
where $B_{2}(n)=2[4k^{2}|\alpha |^{2}\zeta ^{2}+k^{2}(\omega
_{m}t)^{2}(2n+1)]^{-1/2}$ is a normalization coefficient for each state $%
|\psi _{2}(n)\rangle _{\omega _{m}t\ll 1}$ and $\zeta =[(\omega
_{m}t)^{2}\sin \beta ]/2+\omega _{m}t\cos \beta $.

Therefore, this is Eq. (\ref{bb}) in main text.

For Eq. (\ref{yynn}), over all $n$ component, then the final total state of
the pointer is $\rho _{m}^{dis}=(1-z)\sum_{n=0}z^{n}\psi _{2}(n)\rangle
_{\omega _{m}t\ll 1}\langle \psi _{2}(n)|_{\omega _{m}t\ll 1}/B_{2}^{tot}$,
and substituting it into Eq. (\ref{eq:1-5}), then
\begin{equation}
\langle q(t)\rangle _{\omega _{m}t\ll 1}=k^{2}|\alpha |\zeta \omega
_{m}t\sigma _{q}^{2}/(\sigma B_{2}^{tot})  \label{nnnn}
\end{equation}%
where $B_{2}^{tot}=(4k^{2}|\alpha |^{2}\zeta |^{2}\sigma ^{2}+k^{2}(\omega
_{m}t)^{2}\sigma _{q}^{2})/(4\sigma ^{2})$ is a normalized coefficient for $%
\rho _{m}^{dis}$.

Based on Eq. (\ref{nnnn}), we then obtain the maximal positive value $\sigma
_{q}$ or negative value $-\sigma _{q}$ when $2|\alpha |\zeta =\pm \omega
_{m}t\sigma _{q}/\sigma $, respectively. Therefore, the $|\psi
_{2}(n)\rangle $ components corresponding to the maximal positive and
negative amplification, respectively, are $|\psi _{2}(n)\rangle _{\max
,\omega _{m}t\ll 1}=[\sigma _{q}/\sigma \pm (c+c^{\dagger })]|n\rangle /%
\sqrt{2}$ (unnormalized). Then the mirror state achieving the maximal
positive and negative amplification, respectively, are $\rho
_{m}^{dis}=(1-z)\sum_{n=0}z^{n}|\psi _{2}(n)\rangle _{\max ,\omega _{m}t\ll
1}\langle \psi _{2}(n)|_{\max ,\omega _{m}t\ll 1}/4$. It is obvious that the
amplification with displacement thermal state pointer is much larger than
that with pure state pointer \cite{Aharonov88,Simon11,Li14,Li15} since its
maximal value is the ground state fluctuation $\sigma $. Therefore, thermal
noise effect of the pointer (mirror) is beneficial for the amplification of
the mirror's displacement.

\section{Dissipation effect in optomechanical system}

The master equation (\ref{ccc}) in the main text is given by

\begin{eqnarray}
d\rho (t)/dt &=&-i[H,\rho (t)]/\hbar +\gamma _{m}\mathcal{D}[c]/(1-z)  \notag
\\
&+&\gamma _{m}z\mathcal{D}[c^{\dag }]/(1-z),  \label{zzy}
\end{eqnarray}%
where $\mathcal{D}[o]=o\rho (t)o^{\dag }-o^{\dag }o\rho (t)/2-\rho
(t)o^{\dag }o/2$.

For the amplification scheme using a phase shifter $\theta $, at time $t\ll
1 $, if we perform a Taylor expansion about $t=0$ till the second order, the
solution of the master equation is approximately
\begin{equation}
\rho (t)=\rho (0)+td\rho (t)/dt+(t^{2}/2!)d^{2}\rho (t)/dt^{2}.  \label{zzz}
\end{equation}%
When the intial state of the total system is $\rho (0)=|\psi _{i}(\theta
)\rangle \langle \psi _{i}(\theta )|\otimes \rho _{th}(z)$ and after the
postselecting state $|\psi _{p}\rangle $ is performed for the system in Eq. (%
\ref{zzz}) and substituting it into Eq. (\ref{eq:1-5}), by carefully
calculation, we can obtain
\begin{align}
\langle q(t)\rangle _{\omega _{m}t\ll 1}& =[2\sigma _{q}^{2}k\omega
_{m}t\sin \theta +k(\omega _{m}t)^{2}(1-\cos \theta )  \notag \\
& -(1/2)\gamma \sigma _{q}^{2}k(\omega _{m}t)^{2}\sin \theta ]/[2-2\cos
\theta  \notag \\
& +\sigma _{q}^{2}k(\omega _{m}t)^{2}\cos \theta ]/\sigma ,  \label{zzzyy}
\end{align}%
where $\gamma =\gamma _{m}/\omega _{m}.$

This is the average displacement $\langle q(t)\rangle_{\omega_{m}t\ll1}$ of
the mirror after postselection plotted in Figure 4(a) in main text.

For the amplification scheme using the displaced thermal state, at time $%
t\ll 1$, if we perform a Taylor expansion about $t=0$ till the third order,
the solution of the master equation is approximately
\begin{eqnarray}
\rho (t) &=&\rho (0)+td\rho (t)/dt+(t^{2}/2!)d^{2}\rho (t)/dt^{2}  \notag \\
&+&(t^{3}/3!)d^{3}\rho (t)/dt^{3}.  \label{zzzzy}
\end{eqnarray}%
When the initial state of the total system is $\rho (0)=|\psi _{i}\rangle
\langle \psi _{i}|\otimes \rho _{th}(z,\alpha )$ and after the postselecting
state $|\psi _{p}\rangle $ is performed for the system in Eq. (\ref{zzzzy})
and substituting it into Eq. (\ref{eq:1-5}), by carefully calculation, we
can obtain

\begin{align}
&\langle q(t)\rangle _{\omega _{m}t\ll 1} =[3\sigma _{q}^{2}k^{2}(\omega
_{m}t)^{2}|\alpha |\cos \theta /\sigma ^{2}+4k^{2}(\omega _{m}t)^{2}  \notag
\\
& (|\alpha|\cos \theta )^{3}-5\sigma _{q}^{2}\gamma k^{2}(\omega
_{m}t)^{3}|\alpha |\cos \theta /(3\sigma ^{2})-3\gamma k^{2}(\omega
_{m}t)^{3}  \notag \\
& (|\alpha |\cos \theta )^{3}]/[\sigma _{q}^{2}k^{2}(\omega
_{m}t)^{2}/(2\sigma ^{2})+2k^{2}(\omega _{m}t)^{2}(|\alpha |\cos \theta )^{2}
\notag \\
& -\gamma k^{2}(\omega _{m}t)^{3}(|\alpha |\cos \theta )^{2}-\sigma
_{q}^{2}\gamma k^{2}(\omega _{m}t)^{3}/(12\sigma ^{2})]  \notag \\
& -2|\alpha |\cos \theta .
\end{align}

This is the average displacement $\langle q(t)\rangle_{\omega_{m}t\ll1}$ of
the mirror after postselection plotted in Figure 5(a) in main text.

\section{The amplification without postselection in optomechanics}

The time evolution operator of the Hamiltonian (\ref{ee}) in the main text
is given by Eq. (\ref{l}). As shown Fig. 1 in the main text, we use only
single cavity A. When thermal state $\rho _{th}(z)$ is considered as a
pointer in cavity A, and if one photon is weakly coupled with the mirror
using (\ref{l}), it can be found that the mirror will be changed from $\rho
_{th}(z)$ to a displacement thermal state,

\begin{equation}
\rho _{th}(z,\xi )=D(\xi (t))\rho _{th}(z)D^{\dag }(\xi (t)).  \label{mmnn}
\end{equation}%
According to the expression of the displacement
\begin{equation}
\langle q\rangle =Tr(\rho _{th}(z,\xi )\hat{q})-Tr(\rho _{th}(z)\hat{q})
\label{mmnnn}
\end{equation}%
with $\hat{q}=\sigma (c+c)$, the average position displacement of the
pointer without the postselection is given by
\begin{equation}
\langle q\rangle =2k(1-\cos \omega _{m}t)\sigma .  \label{mmmnn}
\end{equation}

However, when displacement thermal state $\rho _{th}(z,\alpha )$ is
considered as a pointer in cavity A, and if one photon is weakly coupled
with the mirror (\ref{l}), it can be found that the mirror will be changed
from $\rho _{th}(z,\alpha )$ to a displacement thermal state,%
\begin{equation}
\rho _{th}(z,\varphi ,\xi )=D(\xi (t))\rho _{th}(z,\varphi )D^{\dag }(\xi
(t)),  \label{m}
\end{equation}%
where $\varphi (t)=\alpha e^{-i\omega _{m}t}$. According to the expression
of the displacement
\begin{equation}
\langle q\rangle =Tr(\rho _{th}(z,\varphi ,\xi )\hat{q})-Tr(\rho
_{th}(z,\varphi )\hat{q})  \label{n}
\end{equation}%
with $\hat{q}=\sigma (c+c)$, the average position displacement of the
pointer without the postselection is the same as Eq. (\ref{mmmnn}).

Fom Eq. (\ref{mmmnn}), it can be seen that the position displacement of the
mirror caused by radiation pressure of one photon can not more than $%
4k\sigma $ for any time $t$. In the literature \cite{Marshall03}, we know
that if the displacement of the mirror can be detected experimentally it
should be not smaller than $\sigma $, implying that the displacement of the
mirror reach strong-coupling limit, so $k=g/\omega _{m}$ can not be bigger
than $0.25$ in weak coupling condition \cite{Marshall03}. When $k=g/\omega
_{m}\leq 0.25$ in weak-coupling regime, the maximal displacement of the
mirror $4k\sigma $ can not be more than $\sigma _{q}$, i.e., thermal
fluctuation of the mirror, therefore the displacement of the mirror caused
by one photon can not be detected.

\section{ Probability $P$}

The overall probability of a single photon (\ref{ddd}) in the main text,
generating the superposition state of $|n\rangle $ and $(c+c^{\dagger
})|n\rangle $, is given by

\begin{equation}
P=(1/4)\int_{0}^{\infty }\kappa \exp (-\kappa t)(\sigma _{q}^{2}|\xi
(t)|^{2}/\sigma ^{2}+\Omega ^{2})dt.  \label{ddmm}
\end{equation}%
where $\Omega =\theta ,\phi (\alpha ,t).$

For the first scheme, $P=\sigma _{q}^{2}k^{2}\omega _{m}^{2}/[2\sigma
^{2}(\kappa ^{2}+\omega _{m}^{2})]+\theta ^{2}/4$, and for the second
scheme, let $|\alpha |=\sigma _{q}/(2\sigma )$ and $\beta =0$, then $%
P=\sigma _{q}k^{2}\omega _{m}^{2}(2\kappa ^{2}+5\omega _{m}^{2})/[2\sigma
(\kappa ^{4}+5\kappa ^{2}\omega _{m}^{2}+4\omega _{m}^{4})]$. Therefore, for
the first scheme, $P$ is approximately $6.94k^{2}$ with $\kappa =1.2\times
10^{4}\omega _{m}$, $\theta =0.005$ and $z=0.999999999$, and for the second
scheme, $P$ is approximately $5k^{2}$ with $\kappa =2\times 10^{4}\omega
_{m} $, $z=0.999999999$.


\begin{thebibliography}{99}
\bibitem{Temperature effect} Quantum optomechanical system usually refers to
a high finesse cavity with a movable mirror where the light in the cavity
can give a force on the mirror \cite{Marquardt09,Aspelmeyer14}. When there
is only one photon in the cavity, the displacement of the mirror caused by
the photon is hard to detect in traditional measurement since it is much
smaller than the spread of the mirror wave packet (quantum fluctuation). Of
course, if the mirror is in thermal state (thermal fluctuation), the
mirror's displacement caused by the photon is even more hard to detect in
traditional measurement.
\bibitem{real amplification} If $\langle \psi _{f}|\psi _{i}\rangle $ is
real and imaginary number, the amplification of weak measurement with
thermal state pointer is given in Appendix B.
\bibitem{Temperature} If the resistance to high temperature materials about
optomechanics is created in the future, the amplification of the
displacement caused by one photon can achieve nanometer category and even
micron category.
\end{thebibliography}
\end{document}